%% file: ShKr2018-article-SIAM.tex
\documentclass[article,onefignum,onetabnum]{siamonline171218}
\usepackage{amsfonts}

\input{ShKr2018-shared-SIAM}

\ifpdf
\hypersetup{
  pdftitle={Time correlation functions of equilibrium and nonequilibrium Langevin dynamics: Derivations and numerics using random numbers},
  pdfauthor={Xiaocheng Shang and Martin Kr\"{o}ger}
}
\fi

\usepackage{bm}
\usepackage{cite}               
\newcommand{\dd}{\mathrm{d}}
\newcommand{\q}{{\bf q}}
\newcommand{\p}{{\bf p}}
\newcommand{\kB}{k_\mathrm{B}}
\graphicspath{{figures/}}

\begin{document}

\maketitle

\begin{abstract}
  We study the time correlation functions of coupled linear Langevin dynamics without and with inertia effects, both analytically and numerically. The model equation represents the physical behavior of a harmonic oscillator in two or three dimensions in the presence of friction, noise, and an external field with both rotational and deformational components. This simple model plays pivotal roles in understanding more complicated processes. The presented analytical solution serves as a test of numerical integration schemes, its derivation is presented in a fashion that allows to be repeated directly in a classroom. While the results in the absence of fields (equilibrium) or confinement (free particle) are omnipresent in the literature, we write down, apparently for the first time, the full nonequilibrium results that may correspond, e.g., to a Hookean dumbbell embedded in a macroscopically homogeneous shear or mixed flow field. We demonstrate how the inertia results reduce to their noninertia counterparts in the nontrivial limit of vanishing mass. While the results are derived using basic integrations over Dirac delta distributions, we mention its relationship with alternative approaches involving (i) Fourier transforms, that seems advantageous only if the measured quantities also reside in Fourier space, and (ii) a Fokker--Planck equation and the moments of the probability distribution. The results, verified by numerical experiments, provide additional means of measuring the performance of numerical methods for such systems. It should be emphasized that this manuscript provides specific details regarding the derivations of the time correlation functions as well as the implementations of various numerical methods, so that it can serve as a standalone piece as part of education in the framework of stochastic differential equations and calculus.
\end{abstract}





\input{ShKr2018-article-SIAM.toc-siamonline-mk}

\section{Introduction}

The efficiency and accuracy of numerical solvers for stochastic differential equations (SDEs), including those that are equivalent to
diffusion-type partial differential equations, is difficult to assess without analytical reference solutions at hand. Only for the simplest linear cases, can transient moments and time correlation functions be calculated analytically. For nonlinear SDEs, analytical solutions are generally not available, nevertheless convergence and stability issues have been discussed~\cite{Higham2002,Rodkina2005,Cao2004}. Here we propose an essentially two--dimensional nontrivial, still linear benchmark problem [Langevin dynamics~\cref{eq:Langevin_Nondim_xy}], inspired by the challenging problem of the dynamics of macromolecules, that is still exactly solvable. It includes inertia effects, which are usually neglected as they pose extra problems and because their physical significance is a priori unclear, or any possible related effects are considered ``small''.

The benchmark equation we are going to consider arises in several different contexts, where linear restoring forces are competing with stochastic noise, in the presence of an external field, while both the absence of either the restoring force or the external field are popular special cases that include, for example, the random walk~\cite{Codling2008,Nelson1982}, diffusion~\cite{Kampen2007,Honerkamp1993,Gardiner2009}, charged atom in an electric field~\cite{Isoda1994}, motion of atoms in the presence of gravitational, centrifugal, chemical potential etc., gradients~\cite{Manning1961}, RNA unfolding via laser tweezers~\cite{Manosas2007}, nanomagnets subjected to magnetic fields and superparamegnetization~\cite{LangevinBook2017}, Brownian oscillators~\cite{LangevinBook2017}, dielectric and magnetic permittivity in dilute solutions of macromolecules~\cite{Hayes1970} or ferrofluids~\cite{Fannin1990}, phoretic forces~\cite{phoretic2006}, vibration and photodesorption of diatomic gases~\cite{Loncaric2016}, and rotational relaxation of molecules trapped in a 3D crystal~\cite{Delgado1987}. Including inertia effects into Brownian dynamics (i.e., the overdamped limit of the Langevin dynamics), where they are usually neglected, can help understand origins of departures from the expected behavior, especially at short times, for tracer nanoparticles experiencing both inertia and stochastic forces, in microrheology, or to explain the occurrence of negative storage moduli~\cite{Yang2006,Mizuno2012,Baiesi2009,Baiesi2010}.

Let us introduce one explicit example from the world of polymer physics, dealing with macromolecules, DNA, actin filaments and the alike, as well as materials, bio\-chem\-ical- and engineering sciences, that is captured by our benchmark problem.
The dynamics of a single flexible polymer dissolved in Newtonian solvent, and flexible polymers confined in melts are both, to a first approximation, well captured by the Brownian motion of a linear chain consisting of a number of identical mass points (or beads), permanently interconnected by harmonic springs, and interacting with the surrounding via white noise~\cite{Doi1996,Oettinger1996}. In that case the harmonic spring results are based upon assumptions, that each partial chain, thought to reside between and terminate at the mass points, behaves as an ideal chain, that can be mapped using Kuhn's approach to a random walk. Assuming Stokes' friction hindering the free motion of the mass points due to frequent collisions with the surrounding medium, the strength of the noise is related to the bead friction coefficient via a fluctuation-dissipation relation. The rheological, viscoelastic properties of polymers are very different from those of simple liquids, and can be studied upon considering a polymer dissolved in a solution that is not at rest, but subjected to a flow gradient. While the precise trajectory of the polymer is unavailable because of the stochastic noise, measurable time correlation functions can be calculated analytically.
Since polymeric systems are often overdamped, the inertia, which is quantified by the mass, is thus typically neglected, which is known as the Rouse model~\cite{Rouse1953,Bird1987II} (i.e., in the form of the Brownian dynamics). However, as pointed out in~\cite{Schieber1988}, the inertia of the chains may be expected to be more important for samples in solvents of extremely low viscosity, e.g., ``supercritical solvents'', due to the fact that the dimensionless mass depends inversely upon the solvent viscosity squared.
Upon introducing normal coordinates~\cite{Doi1988,Doi1996}, the differential equations that need to be solved to treat the complete polymer problem with masses~\cite{Kremer1990}, and for polymers subjected to a macroscopic homogeneous flow field~\cite{Kroeger1993}, are identical to the equations of motion of a harmonic oscillator with a single mass, connected with the origin by a spring.

Inertia effects in the context of microbead rheology~\cite{Schieber2012}, where the spring coefficient $k$ is due to an optical trap, appear to improve the agreement with data for dynamical viscosities at high frequencies~\cite{Willenbacher2007}. The inertia effects are known to be quite irrelevant under most common conditions, but should increase with an increasing size of the microbead and softness of the surrounding material~\cite{Venerus2018}. It has also been demonstrated in~\cite{Hinch1975} that the necessity of including the inertial effects for the study of fluid suspensions. Furthermore, in the context of molecular dynamics, the inclusion of the inertia effects leads to possibilities of designing various thermostats, which are powerful tools for sampling the invariant measure~\cite{Leimkuhler2015b,Allen1989,Frenkel2001}.


This manuscript is organized as follows. We present the model Langevin dynamics, its noninertia special (Brownian) case, and introduce dimensionless quantities in~\cref{sec:Model_Equation} to come up with a dimensionless Langevin dynamics suitable for benchmark tests. In~\cref{sec:Derivations}, we derive the time correlation functions of this equation both without and with inertia effects. In addition to demonstrating that the inertia results reduce to their noninertia counterparts in the limit of vanishing mass, we provide two alternative approaches based on (i) the Fourier transform and (ii) the Fokker--Planck equation to obtain the time correlation functions. We review, in~\cref{sec:Numerical_Methods}, various numerical methods used to solve either Brownian dynamics or Langevin dynamics. The available correlation functions are important measures of dynamical fidelity that numerical integrators should be able to reproduce. \Cref{sec:Numerical_Experiments} presents  numerical experiments in both cases, not only verifying the analytical results but also comparing the performance of those numerical methods. A summary and outlook is given in~\cref{sec:Conclusions}.

\section{The model equation}
\label{sec:Model_Equation}

Consider the linear Langevin dynamics with a single harmonic oscillator of mass $m$ in the presence of a streaming background medium with velocity field $\mathbf{u}$, whose equations of motion for its extension, or end-to-end vector $\q(t)$ is given by
\begin{equation}\label{eq:Langevin}
  m \ddot{\q} = -k\q - \gamma \left( \dot{\q} - \mathbf{u} \right) + \sigma \, \boldsymbol{\eta} \,,
\end{equation}
where a dot denotes a derivative with respect to time $t$, $k$ represents a spring coefficient, and the positive friction coefficient $\gamma$ and noise strength $\sigma$ are related via a fluctuation-dissipation relation
\begin{equation}
\sigma^2=2 \gamma\kB T \,,
\end{equation}
where $\kB$ and $T$ denote the Boltzmann constant and absolute temperature, respectively. The Wiener noise vector $\boldsymbol{\eta}(t)$ with independent components is characterized by
\begin{equation}\label{eq:Wiener}
  \left\langle \boldsymbol{\eta}(t) \right\rangle = {\bf 0} \,, \qquad
  \left\langle \boldsymbol{\eta}(t) \, \boldsymbol{\eta}(t') \right\rangle = {\bf 1} \, \delta(t-t') \,,
\end{equation}
where $\langle \cdot \rangle$ denotes an ensemble average and ${\bf 1}$ is the unity matrix. We impose the initial conditions
$\q(-\infty)={\bf 0}$ and $\dot{\q}(-\infty)={\bf 0}$, when we are interested in time correlation functions such as $\langle {\bf q}(t) \cdot {\bf q}(0)\rangle$ that are unaffected by the precise initial conditions and thus symmetric in $t$ in the absence of the assumed homogeneous streaming velocity field $\mathbf{u} = \bm{\kappa} \cdot \q$. The matrix $\bm{\kappa}$ (transposed macroscopic homogeneous velocity gradient) is arbitrary, traceless for the case of incompressible flow, and can be considered to have nonvanishing components only on its
diagonal, and one of the non-diagonal components, if we choose a suitable coordinate system,
\begin{equation}\label{eq:kappa}
  \bm{\kappa} =
  \left( \begin{array}{ccc}
  \kappa_{xx} & \dot{\gamma} & 0 \\
  0 & \kappa_{yy} & 0 \\
  0 & 0 & \kappa_{zz}
  \end{array} \right) \,.
\end{equation}
In the absence of ${\bf u}$ or for a diagonal $\bm{\kappa}$ tensor characterizing elongational flow,~\cref{eq:Langevin} is identical to three uncoupled equations for three scalar components, each of which describes a one-dimensional linear Langevin dynamics with inertia. In what follows we consider a more general case in which the system is subjected to a mixed flow with shear rate $\dot{\gamma}$. In this case, the equations of~\cref{eq:Langevin} for the components do not decouple anymore, and instead read, with $\q=(x,y,z)$,
\begin{subequations}\label{eq:Langevin_xy}
\begin{align}
  m\ddot{x} &= -k_x x - \gamma \left( \dot{x} - \dot{\gamma}y \right) + \sigma \, \eta_x \,, \label{eq:Langevin_x} \\
  m\ddot{y} &= -k_y y - \gamma \dot{y} + \sigma \, \eta_y \,, \label{eq:Langevin_y}
\end{align}
\end{subequations}
and there is no need to write down an extra equation for the $z$-component, as it remains coupled to neither $x$- nor $y$-components. We have also introduced effective spring coefficients $k_\mu\equiv k - \gamma \kappa_{\mu\mu}, \mu \in \{x, y\}$, to incorporate potential contributions from the diagonal of the $\bm{\kappa}$ tensor. To improve the neatness of the presentation, we are going to introduce appropriate abbreviations below. It also turns out that it would be useful to introduce different abbreviations for both noninertia and inertia cases.

\begin{figure}[tb]
\centering
\includegraphics[scale=0.8]{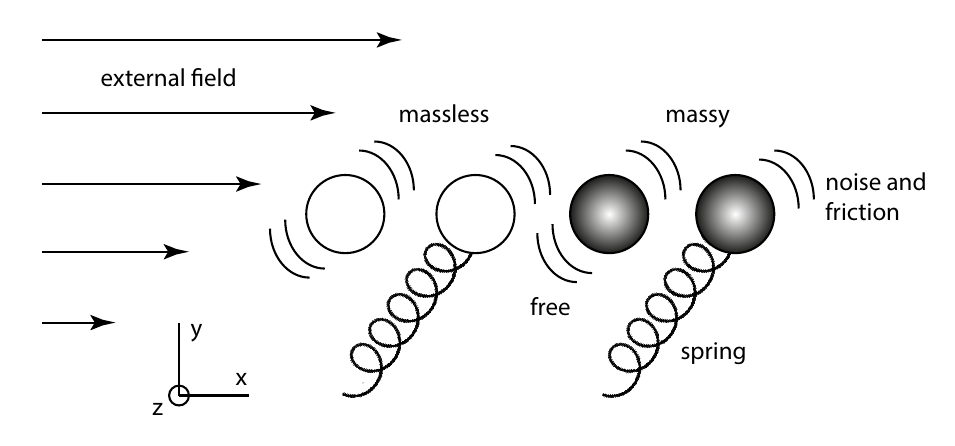}
\caption{\small Schematic descriptions of a variety of possible conditions associated with the Langevin \mbox{dynamics}~\cref{eq:Langevin}. {\bf (a)} free, massless, ideal Brownian {\bf (b)} spring-connected, massless, nonideal Brownian, {\bf (c)} free, inertial, ideal Langevin, and {\bf (d)} spring-connected, inertial, nonideal Langevin cases.}
\label{fig:Schematic}
\end{figure}

For the noninertia ($m=0$) case, associated with Brownian or overdamped Langevin dynamics, we can rewrite~\cref{eq:Langevin_xy} as
\begin{subequations}\label{eq:Langevin_m0_xy}
\begin{align}
  \dot{x} &= \dot{\gamma}y - \omega_x x + \sqrt{2D} \, \eta_x \,, \label{eq:Langevin_m0_x} \\
  \dot{y} &= - \omega_y y + \sqrt{2D} \, \eta_y \,, \label{eq:Langevin_m0_y}
\end{align}
\end{subequations}
having introduced (no summation convention) two characteristic frequencies $\omega_\mu$ and a diffusion coefficient $D$
\begin{equation}\label{eq:Omega_and_D}
  \omega_\mu\equiv \frac{k_\mu}{\gamma} = \frac{k - \gamma \kappa_{\mu\mu}}{\gamma}, \qquad D \equiv \frac{\sigma^2}{2\gamma^2} = \frac{\kB T}{\gamma} \,.
\end{equation}
In fact, we could have eliminated one more parameter by switching to dimensionless time. However, in order to prevent any confusion with the notations, we introduce dimensionless units only for the more advanced inertia case, where dimensionless units pay off more significantly.
To this end we introduce dimensionless position and time for the inertia ($m>0$) case via
\begin{equation}\label{eq:xyt}
  x_{\ast} \equiv \frac{x}{q_\textrm{ref}} \,, \qquad y_{\ast} \equiv \frac{y}{q_\textrm{ref}} \,, \qquad t_{\ast} \equiv \frac{t}{t_\textrm{ref}} \,.
\end{equation}
where reference quantities $q_\textrm{ref}$ and $t_\textrm{ref}$ are chosen as
\begin{equation}\label{eq:reference_quantities}
  q_\textrm{ref} \equiv \frac{\sigma\sqrt{m}}{(\gamma/2)^{3/2}} = \frac{4\sqrt{m\kB T}}{\gamma} \,, \quad t_\textrm{ref} \equiv \frac{2m}{\gamma} \,.
\end{equation}
Upon further introducing dimensionless spring coefficients $s_\mu$ and a dimensionless shear rate $r$ as follows
\begin{equation}\label{eq:sr}
  s_\mu \equiv \frac{4mk_\mu}{\gamma^2} = \frac{4m\omega_\mu}{\gamma} \,, \quad r\equiv \frac{2m\dot{\gamma}}{\gamma} \,,
\end{equation}
the equations of the Langevin dynamics~\cref{eq:Langevin_xy} take the simpler and final form (details in~\Cref{app:Nondimensionalization}), which is our ``benchmark'' problem suitable for analytical and numerical inspections
\begin{subequations}\label{eq:Langevin_Nondim_xy}
\begin{align}
  \ddot{x} &= -s_x x - 2\left(\dot{x}-ry\right) + \eta_x \,, \label{eq:Langevin_Nondim_x} \\
  \ddot{y} &= -s_y y - 2\dot{y} + \eta_y \,, \label{eq:Langevin_Nondim_y}
\end{align}
\end{subequations}
with unaltered~\cref{eq:Wiener} and with only three dimensionless parameters $s_x$, $s_y$, and $r$, representing the strengths of the effective springs (in $x$ and $y$ directions) and the shear rate, respectively. We have omitted all asterisks from~\cref{eq:Langevin_Nondim_xy}, and a dot here denotes a derivative with respect to the reduced time $t_{\ast}=t/t_\textrm{ref}$~\cref{eq:xyt}. All results obtained for the reduced quantities can be converted, according to~\cref{eq:xyt}, to dimensional results involving all six parameters in~\cref{eq:Langevin_xy} by multiplying each $x$, $y$, and $t$ by $q_\textrm{ref}$, $q_\textrm{ref}$, and $t_\textrm{ref}$, respectively. In what follows we derive time correlation functions and other quantities of the linear Langevin dynamics~\cref{eq:Langevin} with~\cref{eq:kappa} under various possible conditions as illustrated in~\cref{fig:Schematic}.

\section{Derivation of time correlation functions}
\label{sec:Derivations}

In this section, we analytically derive time correlation functions of the coupled linear Langevin dynamics~\cref{eq:Langevin} without and with inertia effects.

\subsection{Ideal Brownian dynamics: $m=0, k_x=k_y=0$}
\label{subsec:Brownian_Ideal}

We first consider the ideal Brownian dynamics case where both the inertia and effective springs are absent (i.e., $m=0$ and $k_x=k_y=0$). In this case, the system~\cref{eq:Langevin_m0_xy} describes a freely diffusing massless particle in the presence of a shear flow field and includes classical Brownian motion of a particle in a quiescent background medium as a special case for $\dot{\gamma}=0$. Since the zero'th mode in the normal coordinates~\cite{Doi1988,Doi1996} corresponds to the center of mass of a chain, we indeed need results of the springless case treated here, which are essential for transferring the results of a single harmonic oscillator to those of a bead-spring chain~\cite{Kremer1990,Kroeger1993}, or a dumbbell (see~\cref{sec:dumbbell}). To be more precise, the equations of motion of~\cref{eq:Langevin_m0_xy} in this case reduce to
\begin{subequations}\label{eq:Langevin_m0k0_xy}
\begin{align}
  \dot{x} &= \dot{\gamma}y + \sqrt{2D} \, \eta_x \,, \label{eq:Langevin_m0k0_x} \\
  \dot{y} &= \sqrt{2D} \, \eta_y \,, \label{eq:Langevin_m0k0_y}
\end{align}
\end{subequations}
where $D$ is a diffusion coefficient as confirmed by~\cref{eq:Langevin_m0k0_MSD_yy} below. Since $\left\langle \eta_\mu \right\rangle = 0$, we have $\left\langle \dot{y} \right\rangle = 0$ and $\left\langle \dot{x} \right\rangle = \dot{\gamma} \left\langle y \right\rangle$ on average. Unless otherwise stated, we assume $t\ge 0$ throughout this article, since results associated with $t<0$ can be read off by symmetry arguments. Subjecting to initial conditions of $x(0)=x_0$ and $y(0)=y_0$,~\cref{eq:Langevin_m0k0_xy} are solved by
\begin{subequations}\label{eq:Langevin_m0k0_sol_xy}
\begin{align}
  x(t) - x(0) &= \int_{0}^t \dot{x}(t') \, \dd t' = \int_{0}^t \left[\dot{\gamma}y(t') + \sqrt{2D} \, \eta_x(t') \,\right] \dd t' \,, \label{eq:Langevin_m0k0_sol_x} \\
  y(t) - y(0) &= \int_{0}^t \dot{y}(t') \, \dd t' = \sqrt{2D}\int_{0}^t \, \eta_y(t') \, \dd t' \,. \label{eq:Langevin_m0k0_sol_y}
\end{align}
\end{subequations}
Making use of the properties of the Wiener noise~\cref{eq:Wiener}, we obtain the following two-point time correlation function
\begin{align}\label{eq:Langevin_m0k0_yt1yt2}
  \left\langle [y(t_1)-y_0][y(t_2)-y_0] \right\rangle &= 2D \left\langle \int_{0}^{t_1} \eta_y(t'_1) \, \dd t'_1 \int_{0}^{t_2} \eta_y(t'_2) \, \dd t'_2  \right\rangle \nonumber \\
  &= 2D \int_{0}^{t_1} \int_{0}^{t_2} \left\langle \eta_y(t'_1) \, \eta_y(t'_2) \right\rangle \dd t'_2 \, \dd t'_1  \nonumber \\
  &= 2D \int_{0}^{t_1} \int_{0}^{t_2} \delta(t'_1-t'_2) \, \dd t'_2 \, \dd t'_1 \, \nonumber \\
  &= 2D \int_{0}^{\min(t_1,t_2)} \int_{0}^{\min(t_1,t_2)} \delta(t'_1-t'_2) \, \dd t'_2 \, \dd t'_1  \, \nonumber \\
  &= 2D \int_{0}^{\min(t_1,t_2)} \, \dd t'_1 = 2D \min(t_1,t_2) \,.
\end{align}
The famous mean squared displacement emerges as a special case of~\cref{eq:Langevin_m0k0_yt1yt2} with $t\equiv t_1=t_2$:
\begin{equation}\label{eq:Langevin_m0k0_MSD_yy}
  \left\langle [y(t)-y(0)]^2 \right\rangle = 2Dt \,,
\end{equation}
which actually confirms $D$ to be a diffusion coefficient, as it is usually defined by~\cref{eq:Langevin_m0k0_MSD_yy}. We can further proceed calculating the remaining mean squared displacements (see~\cref{app:moko_xy,app:moko_xx} for proofs)
\begin{equation}
  \left\langle [x(t)-x(0)][y(t)-y(0)] \right\rangle = D \dot{\gamma} t^2 \,,
  \label{3.5}
\end{equation}
and
\begin{equation}\label{eq:Langevin_m0k0_MSD_xx}
  \left\langle \left[x(t)-x(0)\right]^2 \right\rangle = 2D t \left[ 1 + \frac{1}{3} \left( \dot{\gamma} t \right)^2 \right] + \left( \dot{\gamma} y_0 t \right)^2 \,,
\end{equation}
which reduces to the equilibrium result~\cref{eq:Langevin_m0k0_MSD_yy} in the absence of shear (i.e., $\dot{\gamma}=0$). Note that the appearance of the $t^3$ term in~\cref{eq:Langevin_m0k0_MSD_xx} reflects anomalous diffusion that is caused by a velocity change along the flow direction (the $x$--direction) due to the Brownian motion of a particle along the velocity gradient (the $y$--direction), and had been confirmed experimentally in~\cite{Orihara2011,VandenBroeck1982}.

\subsection{Nonideal Brownian dynamics: $m=0, k_x,k_y>0$}
\label{subsec:Brownian_Nonideal}

We next consider the nonideal Brownian dynamics case of the oscillator with effective springs (i.e., $m=0$ and $k_x,k_y>0$), subjected to boundary conditions $x(-\infty)=y(-\infty)=0$. In this case, the system~\cref{eq:Langevin_m0_xy} is formally solved by
\begin{subequations}\label{eq:Langevin_m0k>0_sol_xy}
\begin{align}
  x(t) &= \int_{-\infty}^t \left[ \dot{\gamma}y(t') + \sqrt{2D} \, \eta_x(t') \right] e^{-\omega_x (t-t')} \, \dd t' \,,  \label{eq:Langevin_m0k>0_sol_x} \\
  y(t) &= \sqrt{2D}\int_{-\infty}^t \, \eta_y(t') e^{-\omega_y (t-t')} \, \dd t' \,, \label{eq:Langevin_m0k>0_sol_y}
\end{align}
\end{subequations}
which may be verified by direct insertion. One has $\left\langle x \right\rangle = \left\langle y \right\rangle = 0$ on average. The time correlation function $\left\langle y(t_1)y(t_2) \right\rangle$ can be obtained as (see~\cref{app:mok>o_yt1yt2} for a proof)
\begin{align}\label{eq:Langevin_m0k>0_yt1yt2}
  \left\langle y(t_1)y(t_2) \right\rangle = \frac{D}{\omega_y} e^{-\omega_y |t_1-t_2|} \,,
\end{align}
implying special cases of
\begin{equation}\label{eq:Langevin_m0k>0_yty0}
  \left\langle y(t)y(0) \right\rangle = \frac{D}{\omega_y} e^{-\omega_y t} \,, \quad \left\langle y^2 \right\rangle = \frac{D}{\omega_y} \,.
\end{equation}
The remaining time cross-correlation functions are derived in~\cref{app:mok>o_xty0,app:mok>o_ytx0}
\begin{subequations}
\begin{align}
  \left\langle x(t)y(0) \right\rangle &= D\dot{\gamma}\,
  \frac{(\omega_x+\omega_y)e^{-\omega_y t}-2\omega_y e^{-\omega_x t}}{(\omega_x^2-\omega_y^2)\omega_y} \,, \label{eq:Langevin_m0k>0_xty0} \\
  \left\langle y(t)x(0) \right\rangle &= \frac{D\dot{\gamma}e^{-\omega_y t}}{(\omega_x+\omega_y)\omega_y} \,. \label{eq:Langevin_m0k>0_ytx0}
\end{align}
\end{subequations}
For the stationary mixed moment we thus obtain
\begin{equation}
  \left\langle xy \right\rangle =  \frac{D\dot{\gamma}}{(\omega_x+\omega_y)\omega_y} \,.
\end{equation}
and the autocorrelation in flow $x$--direction becomes, according to~\cref{app:mok>o_xtx0},
\begin{equation}\label{eq:Langevin_m0k>0_xtx0}
  \left\langle x(t)x(0) \right\rangle = \frac{D e^{-\omega_x t}}{\omega_x} + \frac{D \dot{\gamma}^2 \left( \omega_x e^{-\omega_y t}-\omega_y e^{-\omega_x t} \right) }{(\omega_x^2-\omega_y^2)\omega_x\omega_y} \,,
\end{equation}
with the stationary second moment
\begin{equation}
  \left\langle x^2 \right\rangle
  = \frac{D}{\omega_x} + \frac{D\dot{\gamma}^2}{(\omega_x+\omega_y)\omega_x\omega_y} \,.
\end{equation}
In the case of a vanishing shear rate (i.e., $\dot{\gamma}=0$), the system~\cref{eq:Langevin_m0_xy} decouples: both cross-correlations $\langle x(t)y(0) \rangle$~\cref{eq:Langevin_m0k>0_xty0} and $\langle y(t)x(0)\rangle$~\cref{eq:Langevin_m0k>0_ytx0} vanish, and $\langle x(t)x(0) \rangle$~\cref{eq:Langevin_m0k>0_xtx0} reduces to $\langle y(t)y(0) \rangle$~\cref{eq:Langevin_m0k>0_yty0}. Finally, we list the time correlation functions in the special case of pure shear, $\omega\equiv \omega_x=\omega_y$ (i.e., for an oscillator in the absence of elongational flow components), in which neither~\cref{eq:Langevin_m0k>0_xty0} nor~\cref{eq:Langevin_m0k>0_xtx0} diverge:
\begin{subequations}\label{eq:Langevin_m0k>0_benchmark_all}
\begin{align}
  \left\langle y(t)y(0) \right\rangle &= \frac{D e^{-\omega t}}{\omega}  = \frac{\kB T}{k} e^{-kt/\gamma} \,, \label{eq:Langevin_m0k>0_benchmark_yy} \\
  \left\langle x(t)y(0) \right\rangle &= \frac{D \dot{\gamma} \left(1+2\omega t\right) e^{-\omega t} }{2\omega^2} = \frac{\kB T \dot{\gamma}}{2k^2} \left( \gamma + 2kt \right) e^{-k t/\gamma} = \left\langle y(-t)x(0) \right\rangle, \label{eq:Langevin_m0k>0_benchmark_xy} \\
  \left\langle y(t)x(0) \right\rangle &= \frac{D\dot{\gamma}e^{-\omega t}}{2\omega^2} = \frac{\kB T \gamma \dot{\gamma}}{2k^2} e^{-kt/\gamma} = \left\langle x(-t)y(0) \right\rangle, \label{eq:Langevin_m0k>0_benchmark_yx} \\
  \left\langle x(t)x(0) \right\rangle &= \frac{De^{-\omega t}}{\omega} + \frac{D\dot{\gamma}^2 \left(1+\omega t\right) e^{-\omega t}}{2\omega^3} = \frac{\kB T}{2k^3} \left( 2k^2 + \gamma^2\dot{\gamma}^2 + \gamma\dot{\gamma}^2 kt \right) e^{-kt/\gamma} \,. \label{eq:Langevin_m0k>0_benchmark_xx}
\end{align}
\end{subequations}
More specifically, the stationary moments are read off at $t=0$,
\begin{equation}\label{eq:Langevin_m0k>0_benchmark_x2y2}
  \left\langle y^2 \right\rangle = \frac{D}{\omega} = \frac{\kB T}{k} \,, \quad
  \left\langle xy \right\rangle = \frac{D\dot{\gamma}}{2\omega^2} = \frac{\kB T\gamma\dot{\gamma}}{2k^2} \,,  \quad \left\langle x^2 \right\rangle = \frac{D}{\omega} + \frac{D\dot{\gamma}^2}{2\omega^3} = \frac{\kB T}{k} + \frac{\kB T\gamma^2\dot{\gamma}^2}{2k^3} \,.
\end{equation}
We can furthermore derive the mean squared displacement in flow gradient $y$--direction
\begin{equation}\label{eq:Langevin_m0k>0_MSD_yy}
  \left\langle [y(t)-y(0)]^2 \right\rangle = 2 \left\langle y^2 \right\rangle - 2 \left\langle y(t)y(0) \right\rangle = 2Dt + O(t^2) \,,
\end{equation}
which indicates that the mean squared displacement is linear in $t$ only at small times, which qualitatively differs from what we have derived for the noninertia case,~\cref{eq:Langevin_m0k0_MSD_yy}, in~\cref{subsec:Brownian_Ideal}. In the limit of vanishing effective springs, however, the mean squared displacement~\cref{eq:Langevin_m0k>0_MSD_yy} reduces to~\cref{eq:Langevin_m0k0_MSD_yy}, since $k^{-1}[1-\exp(-\alpha k)]=\alpha + O(k)$.

\subsection{Ideal Langevin dynamics: $m>0, k_x=k_y=0$}
\label{subsec:Langevin_Ideal}

We next consider the ideal Langevin dynamics case of a free particle, an oscillator without effective springs (i.e., $m>0$ and $k_x=k_y=0$)~\cite{Foister1980}. In this case, the dimensionless~\cref{eq:Langevin_Nondim_xy} takes the form
\begin{subequations}\label{eq:Langevin_m>0k0_xy}
\begin{align}
  \ddot{x} &= -2(\dot{x} - ry) + \eta_x \,, \label{eq:Langevin_m>0k0_x} \\
  \ddot{y} &= -2\dot{y} + \eta_y \,, \label{eq:Langevin_m>0k0_y}
\end{align}
\end{subequations}
for which one is mostly interested in mean squared displacements rather than time correlation functions, as the latter depend on the initial conditions. In the absence of shear, both components are independent with each other, and only velocities rather than coordinates appear in the equations of motion.
By comparing~\cref{eq:Langevin_m>0k0_xy} with~\cref{eq:Langevin_m0_xy} and~\cref{eq:Langevin_m0k>0_sol_xy}, we have
\begin{subequations}\label{eq:Langevin_m>0k0_sol_xy}
\begin{align}
  \dot{x}(t) &= \int_{-\infty}^t \left[ 2ry(t') + \eta_x(t') \right] e^{-2(t-t')} \, \dd t' \,, \label{eq:Langevin_m>0k0_sol_x} \\
  \dot{y}(t) &= \int_{-\infty}^t \, \eta_y(t') e^{-2(t-t')} \, \dd t' \,, \label{eq:Langevin_m>0k0_sol_y}
\end{align}
\end{subequations}
where $\dot{x}$ and $\dot{y}$ have the interpretation of the velocities. We can read off the velocity autocorrelation function and the mean squared displacement, respectively, from~\cref{eq:Langevin_m0k>0_yt1yt2}--\cref{eq:Langevin_m0k>0_yty0}
upon inspecting the case of $D=1/2$ and $\omega_y=2$ in~\cref{eq:Langevin_m0k>0_sol_y}. This yields
\begin{equation}\label{eq:Langevin_m>0k0_VAF}
  \left\langle \dot{y}(t)\dot{y}(0) \right\rangle = \frac{1}{4} e^{-2t} \,,
\end{equation}
and, as shown in~\cref{app:m>oko_yy},
\begin{equation}\label{eq:Langevin_m>0k0_MSD}
  \left\langle [y(t)-y(0)]^2 \right\rangle = \frac{1}{8} \left( 2t + e^{-2t} - 1 \right) \,.
\end{equation}
Re-dimensionalizing~\cref{eq:Langevin_m>0k0_VAF} the more familiar version of the dimensional velocity autocorrelation function arises
\begin{equation}\label{eq:Langevin_m>0k0_VAF_dim}
  \left\langle \dot{y}(t)\dot{y}(0) \right\rangle = \frac{\kB T}{m} e^{-\gamma t/m } \,.
\end{equation}
In this ideal (free, springless, $k=0$) case, the integrated velocity autocorrelation function turns out to be the diffusion coefficient,
\begin{equation}
  \int_0^{\infty} \left\langle \dot{y}(t)\dot{y}(0) \right\rangle \dd t = D \equiv \frac{\kB T}{\gamma}\,.
\end{equation}
Similarly, re-dimensionalizing~\cref{eq:Langevin_m>0k0_MSD} yields the dimensional mean squared displacement,
\begin{equation}
 \left\langle [y(t)-y(0)]^2 \right\rangle = \frac{2m \kB T}{\gamma^2} \left( \gamma t/m + e^{-\gamma t/m} - 1 \right) = \frac{\kB T}{m} t^2 + O(t^3) \,.
\end{equation}
While this expression is quadratic in $t$ at small times, it reaches $2Dt$ (the diffusive regime) for large times (i.e., $\gamma t/m \gg 1$). A similar calculation, where the boundary condition plays a role as in~\cref{subsec:Brownian_Ideal}, can be performed to obtain the mean squared displacement in $x$--direction.
The mean squared velocity $\langle\dot{y}^2\rangle=\kB T/m$~\cref{eq:Langevin_m>0k0_VAF_dim} is in agreement with the equipartition theorem here, in sharp contrast with Brownian dynamics, for which
$\langle\dot{y}^2\rangle =  2 D \delta(0)$ involves the diverging Dirac delta distribution.

\subsection{Nonideal Langevin dynamics: $m>0, k\equiv k_x=k_y>0$}
\label{subsec:Langevin_Nonideal}

We finally consider the most general nonideal Langevin dynamics case with both inertia and effective springs (i.e., $m>0$ and $k\equiv k_x=k_y>0$). For the sake of simplicity we assume $s \equiv s_x = s_y$ in this case, and the equations of motion of the dimensionless system~\cref{eq:Langevin_Nondim_xy} read,
\begin{subequations}\label{eq:Langevin_m>0k>0_xy}
\begin{align}
  \ddot{x} &= -s x - 2(\dot{x}-ry) + \eta_x \,, \label{eq:Langevin_m>0k>0_x} \\
  \ddot{y} &= -s y - 2\dot{y} +  \eta_y \,. \label{eq:Langevin_m>0k>0_y}
\end{align}
\end{subequations}
As demonstrated in~\cref{app:m>ok>o_y}, the solution of~\cref{eq:Langevin_m>0k>0_y} subjected to initial conditions of $y(-\infty)=0$ and $\dot{y}(-\infty)=0$ appropriate for the calculation of correlation functions is given by
\begin{equation}\label{eq:Langevin_m>0k>0_sol_y}
  y(t) = \frac{1}{2\sqrt{1-s}} \left[ G_y(t,s_-) - G_y(t,s_+) \right],
\end{equation}
where
\begin{equation}\label{eq:Langevin_m>0k>0_sol_Gy}
  G_y(t,s') \equiv \int_{-\infty}^t e^{-s'(t-t')} \, \eta_y(t') \, \dd t' \,,
\end{equation}
with the abbreviation
\begin{equation}\label{eq:Langevin_m>0k>0_sol_spm}
  s_\pm = 1 \pm \sqrt{1-s} \,.
\end{equation}
Similarly, we can also obtain the solution of~\cref{eq:Langevin_m>0k>0_x} as
\begin{equation}\label{eq:Langevin_m>0k>0_sol_x}
  x(t) = \frac{1}{2\sqrt{1-s}} \left[ G_x(t,s_-) - G_x(t,s_+) \right],
\end{equation}
where
\begin{equation}\label{eq:Langevin_m>0k>0_sol_Gx}
  G_x(t,s') \equiv \int_{-\infty}^t e^{-s'(t-t')} \left[ 2ry(t') + \eta_x(t') \right] \dd t' \,.
\end{equation}
Subsequently, we can derive a variety of dimensionless time correlation functions as in~\cref{subsec:Brownian_Nonideal} (details of derivations in~\cref{app:m>ok>o_yty0,app:m>ok>o_xty0,app:m>ok>o_ytx0,app:m>ok>o_xtx0}):
\begin{subequations}\label{eq:Langevin_m>0k>0_benchmark_all}
\begin{align}
  \left\langle y(t)y(0) \right\rangle &\;=\; \frac{C^+_1 + C^-_1}{8 s\sqrt{1-s}} \,, \label{eq:Langevin_m>0k>0_benchmark_yy} \\
  \left\langle x(t)y(0) \right\rangle &\;=\; \frac{r \left( A^+ - A^- \right)}{8s^2(1-s)^{3/2} } = \left\langle y(-t)x(0) \right\rangle, \label{eq:Langevin_m>0k>0_benchmark_xy} \\
  \left\langle y(t)x(0) \right\rangle &\;=\; \frac{r \left( C^-_2 - C^+_2 \right)}{16s^2\sqrt{1-s} } = \left\langle x(-t)y(0) \right\rangle, \label{eq:Langevin_m>0k>0_benchmark_yx} \\
  \left\langle x(t)x(0) \right\rangle &\;=\; \frac{C^+_1 + C^-_1}{8 s\sqrt{1-s}} + \frac{r^2 \left( B^+ + B^- \right)}{16s^3(1-s)^{3/2}} \,, \label{eq:Langevin_m>0k>0_benchmark_xx}
\end{align}
\end{subequations}
with the dimensionless, reduced time-dependent coefficients
\begin{subequations}\label{eq:Langevin_m>0k>0_abbr_ABC}
\begin{align}
  A^\pm &= \left[ 1 + \left( \frac{1}{2} + t \right) (1-s) \pm (2+t)\sqrt{1-s} \right] C_2^\pm \,,  \label{eq:Langevin_m>0k>0_abbr_A} \\
  B^\pm &= \left[ \sqrt{1-s} \left( st + s + 1 \right) \pm \left( 2s - 1 \right) \right] C_2^\pm \,,  \label{eq:Langevin_m>0k>0_abbr_B} \\
  C_n^\pm &= \left( \sqrt{1-s} \mp 1\right)^n \exp\left[ - \left( 1 \pm \sqrt{1-s} \right) t \right]. \label{eq:Langevin_m>0k>0_abbr_C}
\end{align}
\end{subequations}
More specifically, for $t=0$,~\cref{eq:Langevin_m>0k>0_benchmark_all} become
\begin{equation}
  \left\langle y^2 \right\rangle = \frac{1}{4s} \,, \quad
  \left\langle xy \right\rangle = \frac{r}{4s^2} \,, \quad \left\langle x^2 \right\rangle = \frac{1}{4s} + \frac{r^2(s+4)}{8s^3} \,.
\end{equation}
As in~\cref{subsec:Brownian_Nonideal}, in the case of a vanishing shear rate (i.e., $\dot{\gamma}=0$ and subsequently $r=0$), the system~\cref{eq:Langevin_m>0k>0_xy} decouples: both cross correlations $\langle x(t)y(0) \rangle$~\cref{eq:Langevin_m>0k>0_benchmark_xy} and $\langle y(t)x(0)\rangle$~\cref{eq:Langevin_m>0k>0_benchmark_yx} vanish, and $\langle x(t)x(0) \rangle$~\cref{eq:Langevin_m>0k>0_benchmark_xx} reduces to $\langle y(t)y(0) \rangle$~\cref{eq:Langevin_m>0k>0_benchmark_yy}, which can be rewritten as
\begin{equation}\label{eq:Langevin_m>0k>0_benchmark_yy_hyp}
  \left\langle y(t)y(0) \right\rangle = \frac{1}{4s} \left[ \cosh(t\sqrt{1-s}) + \frac{\sinh(t\sqrt{1-s})}{\sqrt{1-s}} \right] e^{-t} \,.
\end{equation}
Re-dimensionalizing~\cref{eq:Langevin_m>0k>0_benchmark_yy_hyp} yields the dimensional time correlation function
\begin{equation}
  \left\langle y(t)y(0) \right\rangle = \frac{\kB T}{k} \left[ \cosh\left(\nu t\right) + \frac{\gamma}{2m\nu} \sinh\left(\nu t\right)\right] e^{-\gamma t/2m} \,,
\end{equation}
where
\begin{equation}
  \nu = \sqrt{\gamma^2/4m^2 - k/m} \,,
\end{equation}
which is in perfect agreement with the dimensional result of~\cite{Wang1945}. More specifically, $\left\langle y^2 \right\rangle$ can be alternatively obtained via the Gibbs--Boltzmann distribution, given $U(y)=ky^2/2$ for the harmonic oscillator,
\begin{equation}
  \left\langle y^2 \right\rangle = \frac{\int_{-\infty}^\infty y^2 \exp[-U(y)/\kB T] \, \dd y}{\int_{-\infty}^\infty \exp[-U(y)/\kB T] \, \dd y} = \frac{\kB T}{k} \,.
\end{equation}
We can furthermore derive the mean squared displacement of
\begin{equation}
  \left\langle [y(t)-y(0)]^2 \right\rangle = 2 \left\langle y^2 \right\rangle - 2 \left\langle y(t)y(0) \right\rangle =
  \frac{\kB T}{m} \left( 1 - \frac{\gamma^2}{4mk} \right) t^2
   + O(t^3) \,,
\end{equation}
which indicates that the mean squared displacement is quadratic in $t$ at small times.

\subsection{Connection between noninertia and inertia results}
\label{subsec:Connections}

To demonstrate that the noninertia results of the time correlation functions in~\cref{subsec:Brownian_Nonideal} are special cases (i.e., in the limit of vanishing mass) of the results with inertia in~\cref{subsec:Langevin_Nonideal}, we have to first write down the time correlation functions~\cref{eq:Langevin_m>0k>0_benchmark_all} using dimensional quantities. To this end we reintroducing the original dimensional variables $m$, $k$, $\gamma$, $\sigma$, $\kB T$, $\dot{\gamma}$, and $t$. This is done by multiplying each time correlation function by $q^2_\textrm{ref}$, and subsequently replacing $t\rightarrow t/t_\textrm{ref}$ and expanding $s$ and $r$ using the definitions in~\cref{eq:sr}. Throughout this subsection $\rightarrow$ stands for ``going from dimensionless to dimensional''. By performing Taylor series expansions in $m$ around $m=0$, we obtain some helpful intermediate results:
\begin{subequations}\label{eq:Taylor_factor_all}
\begin{align}
  \frac{q^2_\textrm{ref}}{8s\sqrt{1-s}} &\;\rightarrow\; \frac{\kB T}{2k\sqrt{1-4mk/\gamma^2}} = \frac{\kB T}{2k} + O(m) \,, \label{eq:Taylor_factor_yy} \\
  \frac{r q^2_\textrm{ref}}{8s^2(1-s)^{3/2}} &\;\rightarrow\; \frac{\kB T \gamma \dot{\gamma}}{4k^2} + O(m) \,, \label{eq:Taylor_factor_xy} \\
  \frac{r q^2_\textrm{ref}}{16 s^2\sqrt{1-s}} &\;\rightarrow\; \frac{\kB T \gamma \dot{\gamma}}{8k^2} + O(m) \,, \label{eq:Taylor_factor_yx} \\
  \frac{r^2 q^2_\textrm{ref}}{16 s^3(1-s)^{3/2}} &\;\rightarrow\; \frac{\kB T \gamma^2 \dot{\gamma}^2}{16k^3} + O(m) \,, \label{eq:Taylor_factor_xx}
\end{align}
\end{subequations}
as well as
\begin{subequations}\label{eq:Taylor_factor_spm_all}
\begin{align}
  \left( \mp s_\mp \right)^n &\rightarrow \left( 1 \mp 1 \right)^n - \frac{2n(1\mp 1)^{n-1}}{\gamma^2/mk} \pm \frac{2n(1 \mp 1)^{n-2} [1 \pm (n\!-\!2)]}{(\gamma^2/mk)^2} + O(m^3)
  \,, \label{eq:Taylor_mpsmpn} \\
  s_\pm t &\rightarrow \frac{\gamma t}{2m} \left( 1 \pm \sqrt{1-4mk/\gamma^2} \right)
  = \mp \frac{kt}{\gamma} \left( 1 + \frac{mk}{\gamma^2} \right) + \left( 1 \pm 1 \right) \frac{\gamma t}{2m} + O(m^2), \label{eq:Taylor_factor_spmt}
\end{align}
\end{subequations}
where $t$ on the left-hand side in~\cref{eq:Taylor_factor_spmt} is the dimensionless time, whereas $t$ on the right-hand side denotes the dimensional time. For small $m$ (and $n>0$),~\cref{eq:Taylor_factor_spm_all} implies
\begin{subequations}\label{eq:Taylor_factor_add_all}
\begin{align}
  (+s_+)^n  &\;\rightarrow\; 2^n + O(m) \,,\label{eq:Taylor_factor_pspn} \\
  (-s_-)^n  &\;\rightarrow\; O(m^n) \,, \label{eq:Taylor_factor_msmn} \\
  e^{-s_+t} &\;\rightarrow\; e^{-\gamma t/m} \,, \label{eq:Taylor_factor_emspt} \\
  e^{-s_-t} &\;\rightarrow\; e^{-kt/\gamma} + O(m)\, \label{eq:Taylor_factor_emsmt}
\end{align}
\end{subequations}
where we kept $\exp(-\gamma t/m)$ as it cannot be Taylor expanded; it asymptotically vanishes in the limit $m \rightarrow 0$ as long as $\gamma t>0$. We recall from~\cref{eq:Langevin_m>0k>0_abbr_C} that the coefficients $C_n^\pm$ are given by $C^\pm_n = (\mp s_\mp)^n e^{-s_\pm t}$. With the help of~\cref{eq:Taylor_factor_add_all} we find
\begin{subequations}
\begin{align}
 C^+_n = (-s_-)^n e^{-s_+ t} &\;\;\rightarrow\;\; O(m^n)e^{-\gamma t/m} \,, \label{eq:Taylor_factor_Cp} \\
 C^-_n = (+s_+)^n e^{-s_- t} &\;\;\rightarrow\;\; 2^n e^{-kt/\gamma} + O(m) \,,
 \label{eq:Taylor_factor_Cm}
\end{align}
\end{subequations}
and thus only the coefficients $C_n^-$ survive in the limit of vanishing $m$,
\begin{subequations}
\begin{align}
 \lim_{m\rightarrow 0} \left\langle y(t)y(0) \right\rangle &\;\rightarrow\;
 \lim_{m\rightarrow 0} \frac{q_\textrm{ref}^2(C^+_1 + C^-_1)}{8s\sqrt{1-s}} = \frac{\kB T}{k} e^{-kt/\gamma} \,, \label{3.42a}\\
 \lim_{m\rightarrow 0} \left\langle y(t)x(0) \right\rangle &\;\rightarrow\;
 \lim_{m\rightarrow 0} \frac{q_\textrm{ref}^2r(C^-_2 - C^+_2)}{16s^2\sqrt{1-s}} = \frac{\kB T \gamma \dot{\gamma}}{2k^2} e^{-kt/\gamma} \,,\label{3.42b}
\end{align}
\end{subequations}
where~\cref{eq:Taylor_factor_yy,eq:Taylor_factor_yx} have been used.
\Cref{3.42a,3.42b} coincide with the results~\cref{eq:Langevin_m0k>0_benchmark_yy,eq:Langevin_m0k>0_benchmark_yx} obtained by a direct calculation with $m=0$. To calculate the remaining two correlations, we begin with two intermediate results that both follow from~\cref{eq:Langevin_m>0k>0_abbr_ABC},
\begin{subequations}
\begin{align}
  \frac{A^\pm}{C_2^\pm} \;\;&\rightarrow\;\; 1 + \left( \frac{1}{2} + \frac{\gamma t}{2m} \right) \left( 1 - \frac{4mk}{\gamma^2} \right)
  \pm \left( 2 + \frac{\gamma t}{2m} \right) \sqrt{1-\frac{4mk}{\gamma^2}} \nonumber \\
  &=\;\;\left( 1 \pm 1 \right) \frac{\gamma t}{2m} + \frac{3}{2} \pm 2 - \left( 2 \pm 1 \right) \frac{kt}{\gamma} + O(m) \,, \label{eq:Taylor_factor_AdC} \\
  \frac{B^\pm}{C_2^\pm} \;\;&\rightarrow\;\; \sqrt{1-\frac{4mk}{\gamma^2}} \left( \frac{2kt}{\gamma} + \frac{4mk}{\gamma^2} + 1 \right) \pm \left( \frac{8mk}{\gamma^2} - 1 \right) \nonumber \\
  &=\;\; \left( 1 \mp 1 \right) + \frac{2kt}{\gamma} + O(m) \,.
  \label{eq:Taylor_factor_BdC}
\end{align}
\end{subequations}
Since $m^{-1} C_n^+$ vanishes according to~\cref{eq:Taylor_factor_Cp} as $O(m^{n-1})e^{-\gamma t/m}$, both $A^+$ and $B^+$ vanish in the limit of vanishing mass, and the remaining $A^-$ and $B^-$ are
\begin{subequations}
\begin{align}
 A^- &\;\rightarrow\; \left( \frac{3}{2}-2 - \frac{kt}{\gamma}\right) 2^2 e^{-kt/\gamma} + O(m) \,, \\
 B^- &\;\rightarrow\; \left( 2 + \frac{2kt}{\gamma}\right) 2^2 e^{-kt/\gamma} + O(m) \,,
\end{align}
\end{subequations}
such that we find ourselves, with the help of~\cref{eq:Taylor_factor_xy},~\cref{eq:Taylor_factor_xx},~\cref{eq:Taylor_factor_AdC}, and~\cref{eq:Taylor_factor_BdC},
\begin{subequations}
\begin{align}
  \lim_{m\rightarrow 0} \left\langle x(t)y(0) \right\rangle
  &\;\rightarrow\; \lim_{m\rightarrow 0} \frac{q^2_\textrm{ref} r \left( A^+ - A^- \right)}{8 s^2(1-s)^{3/2}} = \frac{\kB T \dot{\gamma}}{2k^2} \left( \gamma + 2kt \right) e^{-kt/\gamma} \,,
  \\
  \lim_{m\rightarrow 0} \left\langle x(t)x(0) \right\rangle &\;\rightarrow\; \lim_{m\rightarrow 0} \left\langle y(t)y(0) \right\rangle + \lim_{m\rightarrow 0} \frac{q^2_\textrm{ref} r^2 \left( B^+ + B^- \right)}{16 s^3(1-s)^{3/2}} \nonumber \\
  & \quad = \frac{\kB T}{2k^3} \left( 2k^2 + \gamma^2\dot{\gamma}^2 + \gamma \dot{\gamma}^2 kt \right) e^{-kt/\gamma} \,,
\end{align}
\end{subequations}
in complete agreement with the results obtained by the direct calculation with $m=0$,~\cref{eq:Langevin_m0k>0_benchmark_xy} and~\cref{eq:Langevin_m0k>0_benchmark_xx}, respectively.

\subsection{Alternative approach via Fourier transform}
\label{subsec:Alternative}

We have demonstrated in~\cref{subsec:Langevin_Nonideal} how the time correlation functions for the most general nonideal Langevin dynamics case can be derived via a direct approach, where the Dirac delta distribution is eliminated by integrating over it. In this section, we outline an alternative approach utilizing Fourier transforms, which relates to the Wiener--Khinchin theorem. In this case, we eliminate the Dirac delta distribution by noting that $\delta(t)$ is the inverse Fourier-transformed ``one'' (see~\cref{eq:delta}). In what follows, we only demonstrate how this alternative approach works in an example of the time correlation function of $\left\langle y(t)y(0) \right\rangle$~\cref{eq:Langevin_m>0k>0_benchmark_yy}. Upon substituting $t-t'$ by $t_1$, we can rewrite~\cref{eq:Langevin_m>0k>0_sol_y} more conveniently as
\begin{align}\label{eq:Langevin_m>0k>0_sol_y_alt}
  y(t) &= \frac{1}{\sqrt{1-s}} \int_{-\infty}^t \sinh \left[ (t-t')\sqrt{1-s} \right] e^{-\left(t-t'\right)} \, \eta_y(t') \, \dd t'
  \nonumber \\ 
  &= \frac{1}{\sqrt{1-s}} \int_0^\infty \sinh \left[ t_1\sqrt{1-s} \right] e^{-t_1} \, \eta_y(t-t_1) \, \dd t_1
  \nonumber \\
  &= \int_0^\infty \Omega_{t_1} \, \eta_y(t-t_1) \, \dd t_1 \,,
\end{align}
with a weighting function $\Omega$ defined as
\begin{align}\label{eq:Omega}
  \Omega_\nu &\equiv \frac{e^{-\nu} \sinh(\nu\sqrt{1-s})}{\sqrt{1-s}}
  = \left\{\begin{array}{ll}
  (1-s)^{-1/2} e^{-\nu} \sinh(\nu\sqrt{1-s}) \,, & s\le 1 \, ; \\[1mm]
  (s-1)^{-1/2} e^{-\nu} \sin(\nu\sqrt{s-1}) \,, & s>1 \,,
  \end{array}\right.
\end{align}
where we have also mentioned the purely real-valued version for $s>1$. Now making use of the Fourier transform
\begin{equation}
  \textrm{FT}\{f(p)\}(t) = \frac{1}{\sqrt{2\pi}} \int_{-\infty}^\infty f(p) e^{ipt} \, \dd p \,,
\end{equation}
as well as the basic identity
\begin{equation}\label{eq:delta}
 \delta(t) = \frac{1}{2\pi} \int_{-\infty}^{\infty} e^{ipt} \, \dd p = \frac{1}{\sqrt{2\pi}}\, \textrm{FT}\left\{1\right\}(t) \,,
\end{equation}
the time correlation function of $\left\langle y(t)y(0) \right\rangle$~\cref{eq:Langevin_m>0k>0_benchmark_yy} can be recalculated as follows:
\begin{align}
  \langle y(t)y(0)\rangle &= \int_0^\infty \int_0^\infty \Omega_{t_1} \Omega_{t_2} \left\langle \eta_y(t-t_1)\eta_y(0-t_2) \right\rangle \dd t_2 \, \dd t_1
  \nonumber \\
  &= \int_0^\infty \int_0^\infty \Omega_{t_1}\Omega_{t_2} \delta(t-t_1+t_2) \, \dd t_2 \, \dd t_1
  \nonumber \\
  &= \frac{1}{2\pi} \int_{-\infty}^{\infty}
  \left[ \int_0^\infty \int_0^\infty \Omega_{t_1}\Omega_{t_2} e^{ip(t-t_1+t_2)} \, \dd t_2  \, \dd t_1 \right] \dd p
  \nonumber \\
  &= \frac{1}{2\pi} \int_{-\infty}^{\infty}
  \frac{e^{ipt}}{p^4-2p^2(s-2)+s^2} \, \dd p
  \nonumber \\
  &= \frac{1}{\sqrt{2\pi}} \,\textrm{FT} \left\{\frac{1}{p^4-2p^2(s-2)+s^2} \right\}(t)
  \nonumber \\
  &= \frac{1}{4s} \left[ \cosh(t\sqrt{1-s}) + \frac{\sinh(t\sqrt{1-s})}{\sqrt{1-s}} \right] e^{-t}
  = \frac{C^+_1 + C^-_1}{8s\sqrt{1-s}} \,.
\end{align}
The remaining time correlation functions in~\cref{subsec:Langevin_Nonideal} can be similarly obtained, although the calculations are more involved.

\subsection{Alternative approach via Fokker--Planck equation}
\label{sec:FokkerPlanck}

A complementary approach to the moments and correlation functions is based on the equivalence between the Langevin dynamics for stochastic variables ${\bf Q}(t)$ and a Fokker--Planck equation for the probability distribution function $f({\bf Q},t)$. The Fokker--Planck equation corresponding to the Langevin dynamics in its rather general form is as follows
\begin{equation}\label{eq:generalQ}
 \dot{\bf Q}={\bf a} + \frac{1}{2}\nabla\cdot{\bf D} + {\bf B}\cdot\bm{\eta} \,, \qquad \nabla = \frac{\partial}{\partial{\bf Q}} \,,
\end{equation}
with ${\bf Q}$ and $t$-dependent vector ${\bf a}$, matrices ${\bf B}$ and ${\bf D}={\bf B}\cdot{\bf B}^T$ fulfills the Fokker--Planck equation
\begin{equation}\label{eq:generalFP}
 \frac{\partial f}{\partial t} = -\nabla \cdot \left( {\bf a} f \right)
  + \frac{1}{2} \nabla \cdot \left( {\bf D} \cdot \nabla f \right) \,.
\end{equation}
In view of~\cref{eq:Langevin_m>0k>0_qp} the benchmark Langevin dynamics~\cref{eq:Langevin_Nondim_xy} is of the form~\cref{eq:generalQ} with ${\bf a}=-{\bf A}\cdot{\bf Q}$ and constant matrices ${\bf A}$ and ${\bf B}$,
\begin{equation}
 {\bf A} =
 \left( \begin{array}{cccc} 0 & 0 & -1 & 0 \\ 0 & 0 & 0 & -1 \\
  s_x & -2r & 2 & 0 \\ 0 & s_y & 0 & 2 \end{array} \right) \,,
 \qquad {\bf B} =
 \left( \begin{array}{cc} {\bf 0} & {\bf 0} \\ {\bf 0} & {\bf 1} \end{array} \right)\,,
 \label{matrixA}
\end{equation}
while ${\bf Q}$ is the four-dimensional vector $({\bf q}, {\bf p}=m\dot{\bf q})$. With ${\bf Y}(t) = \exp[-{\bf A}t]$ the time evolution of the mean value is $\langle{\bf Q}\rangle(t)={\bf Y}\cdot{\bf Q}_0$ and the variance $\bm{\Sigma}=\langle {\bf QQ}\rangle -\langle{\bf Q}\rangle\langle{\bf Q}\rangle$ fulfills~\cite{GardinerBook}
\begin{equation}\label{eq:generalSigma}
 \dot{\bm{\Sigma}} = - \left[ {\bf A} \cdot \bm{\Sigma} + \bm{\Sigma} \cdot {\bf A}^\textrm{T} \right] + {\bf D} \,.
\end{equation}
With $\bm{\Sigma}(t)$ at hand the solution of the Fokker--Planck equation~\cref{eq:generalFP} reads
\begin{equation}
 p({\bf Q},t) = \frac{1}{(2\pi) \sqrt{|\bm{\Sigma}(t)| } } \exp \left\{-\frac{1}{2} \left[ {\bf Q} - \langle {\bf Q} \rangle(t) \right] \cdot \bm{\Sigma}^{-1}(t) \cdot \left[ {\bf Q} - \langle {\bf Q} \rangle(t) \right] \right\},
\end{equation}
and a stationary solution exists only if~\cref{eq:generalSigma} has a solution for $\dot{\bm{\Sigma}}={\bf 0}$, denoted by $\bm{\Sigma}_\infty$. For the special case $s \equiv s_x=s_y$ considered earlier in~\cref{subsec:Langevin_Nonideal}, the eigenvalues of ${\bf A}$ are $s_\pm$ (both twice degenerated), and the eigenvectors are $(-s_+/s,0,1,0)$, ${\bf 0}$, $(-s_-/s,0,1,0)$, and ${\bf 0}$, respectively. The eigenvalues are real-valued and semipositive for $s \in [0,1]$, and become complex-valued for $s>1$. For $t>0$,
\begin{subequations}
\begin{align}
 \langle {\bf Q}(t){\bf Q}(0) \rangle_\textrm{stat} &= e^{-{\bf A} t} \cdot \bm{\Sigma}_\infty \,, \label{eq:generalQtQ0} \\
 \langle {\bf Q}(0){\bf Q}(t) \rangle_\textrm{stat} &= \bm{\Sigma}_\infty \cdot e^{-{\bf A} t} \,.
\end{align}
\end{subequations}
For the spectral density~\cite{GardinerBook}
\begin{equation}
 {\bf S}(\omega) = \int_{-\infty}^{\infty} \langle {\bf Q}(t+\tau){\bf Q}(t) \rangle_\textrm{stat} e^{-i\omega t} \, \dd \tau
 = \left( {\bf A}+i\omega {\bf 1} \right)^{-1} \cdot {\bf D} \cdot \left( {\bf A}^\textrm{T} - i\omega {\bf 1} \right)^{-1} \,,
\end{equation}
the situation is particularly simple, as it involves only ${\bf A}$ and ${\bf D}$, but not $\bm{\Sigma}_\infty$. Solving the linear system of equations~\cref{eq:generalSigma} for $\bm{\Sigma}_\infty$ we obtain
\begin{equation}
 \bm{\Sigma}_\infty = \begin{pmatrix}
 \frac{2s^2+(4+s)r^2}{8s^3} & \frac{r}{4s^2} & 0 & -\frac{r}{8s} \\
 \frac{r}{4s^2} & \frac{1}{4s} & \frac{r}{8s} & 0 \\
 0 & \frac{r}{8s} & \frac{1}{8}\left(2+\frac{r^2}{s}\right) & 0 \\
 -\frac{r}{8s} & 0 & 0 & \frac{1}{4}
 \end{pmatrix}
\end{equation}
and together with the eigensystem of ${\bf A}$ we have verified that~\cref{eq:generalQtQ0} agrees with~\cref{eq:Langevin_m>0k>0_benchmark_yy}.

\subsection{Connections with the dumbbell model}\label{sec:dumbbell}

The so called dumbbell model, where two masses $m$ are connected by a spring with a spring coefficient $k$, is the simplest model to describe the behavior of a drastically coarse-grained polymer molecule, whose equations of motion (subjected to shear with rate $\dot{\gamma}$ and/or elongational flow whose rates are captured by anisotropic spring coefficients $k_x$ and $k_y$) read
\begin{subequations}\label{eq:Dumbbell_x12}
\begin{align}
 m \ddot{x}_1 &=  - k_x \left( x_1 - x_2 \right) - \gamma \left( \dot{x}_1 - \dot{\gamma}y_1 \right) + \sigma \eta_{x_1} \,, \label{eq:Dumbbell_x1} \\
 m \ddot{x}_2 &=  - k_x \left( x_2 - x_1 \right) - \gamma \left( \dot{x}_2 - \dot{\gamma}y_2 \right) + \sigma \eta_{x_2} \,, \label{eq:Dumbbell_x2} \\
 m \ddot{y}_1 &=  - k_y \left( y_1 - y_2 \right) - \gamma \dot{y}_1 + \sigma \eta_{y_1} \,, \label{eq:Dumbbell_y1} \\
 m \ddot{y}_2 &=  - k_y \left( y_2 - y_1 \right) - \gamma \dot{y}_2 + \sigma \eta_{y_2} \,. \label{eq:Dumbbell_y2}
\end{align}
\end{subequations}
Introducing relative (end-to-end) vector components $X = x_2 - x_1$, $Y = y_2 - y_1$, center of mass coordinates $C_x=(x_1+x_2)/2$, $C_y=(y_1+y_2)/2$, and noting that $\sqrt{2}\,\eta_x = \eta_{x_1} \pm \eta_{x_2}$
\cref{eq:Dumbbell_x12} becomes
\begin{subequations}
\begin{align}
 m \ddot{C}_x &= - \gamma \left( \dot{C}_x - \dot{\gamma} C_y \right) + \frac{\sigma}{\sqrt{2}} \eta_x \,, \\
 m \ddot{C}_y &= - \gamma \dot{C}_y + \frac{\sigma}{\sqrt{2}} \eta_y \,,\\
 m \ddot{X} &= - 2 k_x X - \gamma \left( \dot{X} - \dot{\gamma} Y \right)  +  \sqrt{2} \sigma \eta_x \,, \\
 m \ddot{Y} &= - 2 k_y Y - \gamma \dot{Y} +  \sqrt{2} \sigma \eta_y \,.
\end{align}
\end{subequations}
These two uncoupled sets of equations for $X,Y$ and $C_x,C_y$ are of the form studied in~\cref{subsec:Langevin_Ideal,subsec:Langevin_Nonideal}, respectively. With the new 1-variables $k^1_\mu = 2 k_\mu$, $\gamma_1 = \gamma$, and $\sigma^2_1 = 2\sigma^2 = 4\gamma\kB T = 2\gamma_1 \kB T_1$ the end-to-end vector of the elastic dumbbell behaves like a harmonic oscillator with mass $m$, unchanged friction coefficient $\gamma$, but modified spring coefficient $k_\mu^1=2 k_\mu$ and temperature $T_1=2T$.
Therefore, the time correlation functions for the end-to-end vector ${\bf q}$ of the dumbbell model are identical with those obtained for the nonideal cases upon replacing $T$ by $2T$ and $k_\mu$ by $2k_\mu$. Similarly, the dynamics of the center of mass of the dumbbell is captured by the results for the ideal (springless) cases upon replacing $T$ by $T/2$. The overdamped (noninertia) cases of the dumbbell were thus also treated in~\cref{subsec:Brownian_Ideal,subsec:Brownian_Nonideal}.

\section{Numerical methods}
\label{sec:Numerical_Methods}

In this section, we describe numerical methods used to simulate the linear Langevin dynamics~\cref{eq:Langevin} in both noninertia and inertia cases.

\subsection{Brownian dynamics}
\label{subsec:Brownian_Dynamics}

We consider the linear Langevin dynamics with effective springs but without inertia described in~\cref{subsec:Brownian_Nonideal} (i.e.,~\cref{eq:Langevin_m0_xy}), which is also known as the Brownian dynamics.
\begin{equation}\label{eq:Brownian}
  \dot{\q} = - k \q /\gamma + \mathbf{u} + \sqrt{2D} \, \boldsymbol{\eta} \,,
\end{equation}
where $\mathbf{u} = \bm{\kappa} \cdot \q$ is the streaming velocity field with $\bm{\kappa}$ being defined in~\cref{eq:kappa}.

\subsubsection{The Euler--Maruyama (EM) method}

A simple and popular numerical method for a system of stochastic differential equations is the Euler--Maruyama (EM) method, which reads
\begin{equation}\label{eq:EM}
  \q^{n+1} = \q^{n} - h k \q^{n} /\gamma + h \mathbf{u}^{n} + \sqrt{2Dh} \mathbf{R}^{n} \,,
\end{equation}
where $h$ denotes the integration timestep, and $\mathbf{R}^{n}$, resampled at each step, is a vector of independent Gaussian white noise with zero mean and unit variance.

\subsubsection{The limit method}

A simple modification of the Euler--Maruyama method~\cref{eq:EM} leads to the limit method~\cite{Leimkuhler2013}:
\begin{equation}\label{eq:Limit}
  \q^{n+1} = \q^{n} - h k \q^{n} /\gamma + h \mathbf{u}^{n} + \sqrt{Dh/2} \left( \mathbf{R}^{n} + \mathbf{R}^{n+1} \right) \,,
\end{equation}
where $\mathbf{R}^{n}$ and $\mathbf{R}^{n+1}$ are vectors of independent Gaussian white noise with zero mean and unit variance, and it should be noted that $\mathbf{R}^{n+1}$ will become $\mathbf{R}^{n}$ in the subsequent step. It has been showed that such a simple modification could lead to an extra order of weak convergence~\cite{Leimkuhler2014} as well as substantial improvements in sampling accuracy~\cite{Leimkuhler2013}. Note that although the limit method was first derived from the BAOAB method introduced in~\cref{subsec:BAOAB} in the large friction limit~\cite{Leimkuhler2013}, it can also be obtained via a approach of postprocessed integrators~\cite{Vilmart2015}.

\subsection{Langevin dynamics}
\label{subsec:Langevin_Dynamics}

We also consider the most general case of the linear Langevin dynamics with both inertia and effective springs described in~\cref{subsec:Langevin_Nonideal}. Rewriting~\cref{eq:Langevin_m>0k>0_xy} in a more general and first order form yields
\begin{subequations}\label{eq:Langevin_m>0k>0_qp}
\begin{align}
  \dot{\q} &= \p \,, \label{eq:Langevin_m>0k>0_q} \\
  \dot{\p} &= - s \q - 2 \left( \p - \mathbf{u} \right) + \bm{\eta} \,, \label{eq:Langevin_m>0k>0_p}
\end{align}
\end{subequations}
where $\p$ has the interpretation of the momentum, and~\cref{eq:Langevin_m>0k>0_qp} can be considered as the adimensional version of~\cref{eq:Langevin}, using the reference quantities~\cref{eq:reference_quantities} and dimensionless parameters~\cref{eq:sr}.

\subsubsection{The stochastic velocity Verlet (SVV) method}

Building on the popular Verlet method in molecular dynamics and also due to its ease of implementation, the stochastic velocity Verlet (SVV) method~\cite{Melchionna2007} is a popular scheme for Langevin dynamics, whose integration steps read
\begin{subequations}
\begin{align}
    \p^{n+1/2} &= \p^{n} - h s \q^{n}/2 - h \left( \p^{n} - \mathbf{u}^{n} \right) + \sqrt{h/2} \mathbf{R}^{n} \,, \\
    \q^{n+1} &= \q^{n} + h \p^{n+1/2} \,, \\
    \p^{n+1} &= \p^{n+1/2} - h s \q^{n+1}/2 - h \left( \p^{n+1/2} - \mathbf{u}^{n+1} \right) + \sqrt{h/2} \mathbf{R}^{n+1/2} \,,
\end{align}
\end{subequations}
where $\mathbf{R}^{n}$ and $\mathbf{R}^{n+1/2}$, resampled at each step, are vectors of independent Gaussian white noise with zero mean and unit variance.

\begin{figure}[tbh]
\centering
\includegraphics[scale=0.4]{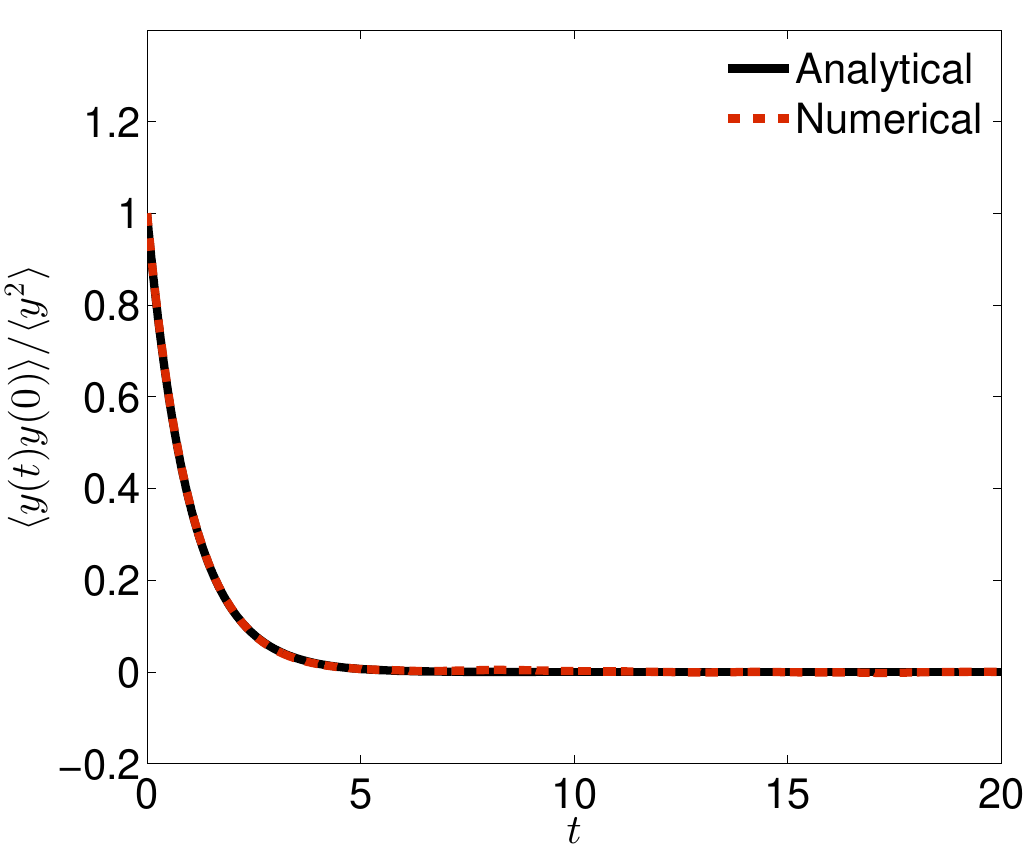}
\includegraphics[scale=0.4]{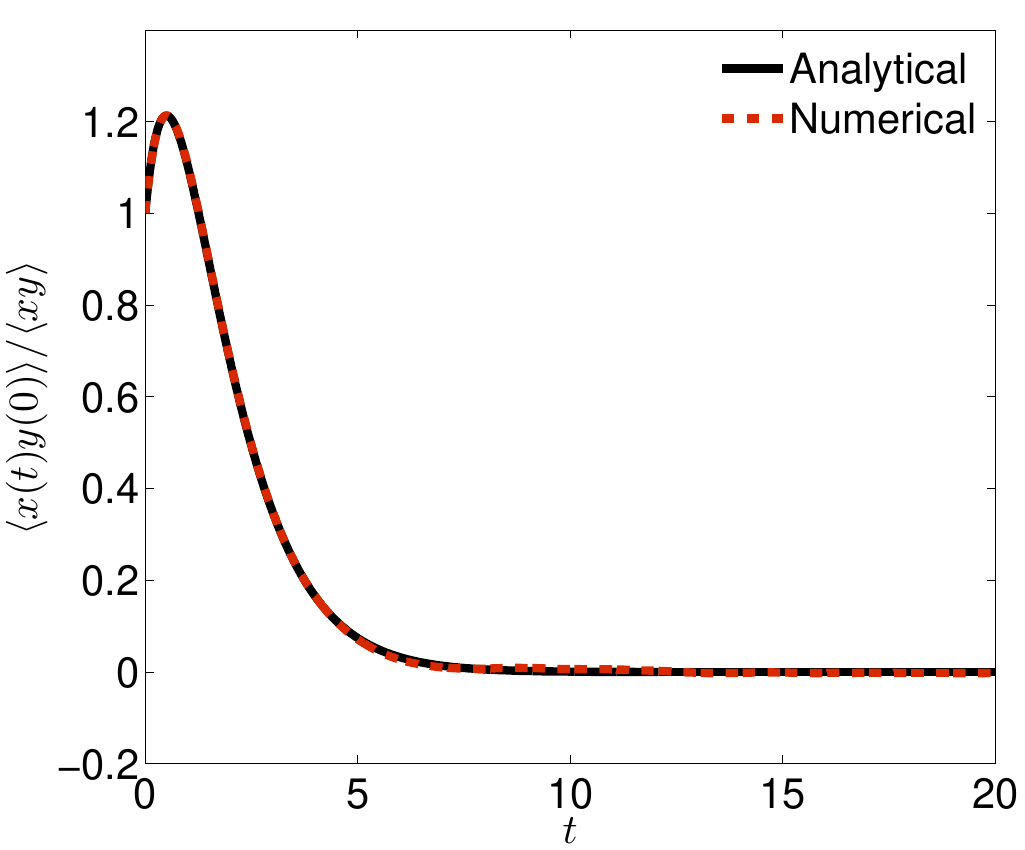}
\includegraphics[scale=0.4]{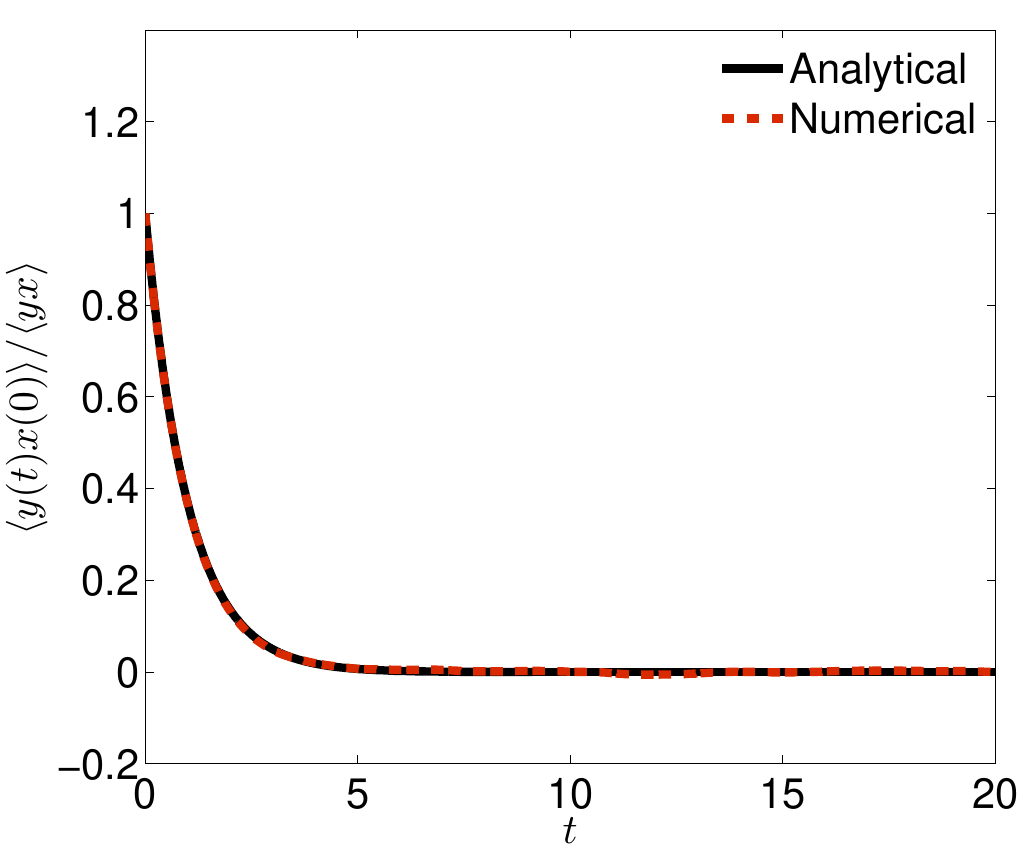}
\includegraphics[scale=0.4]{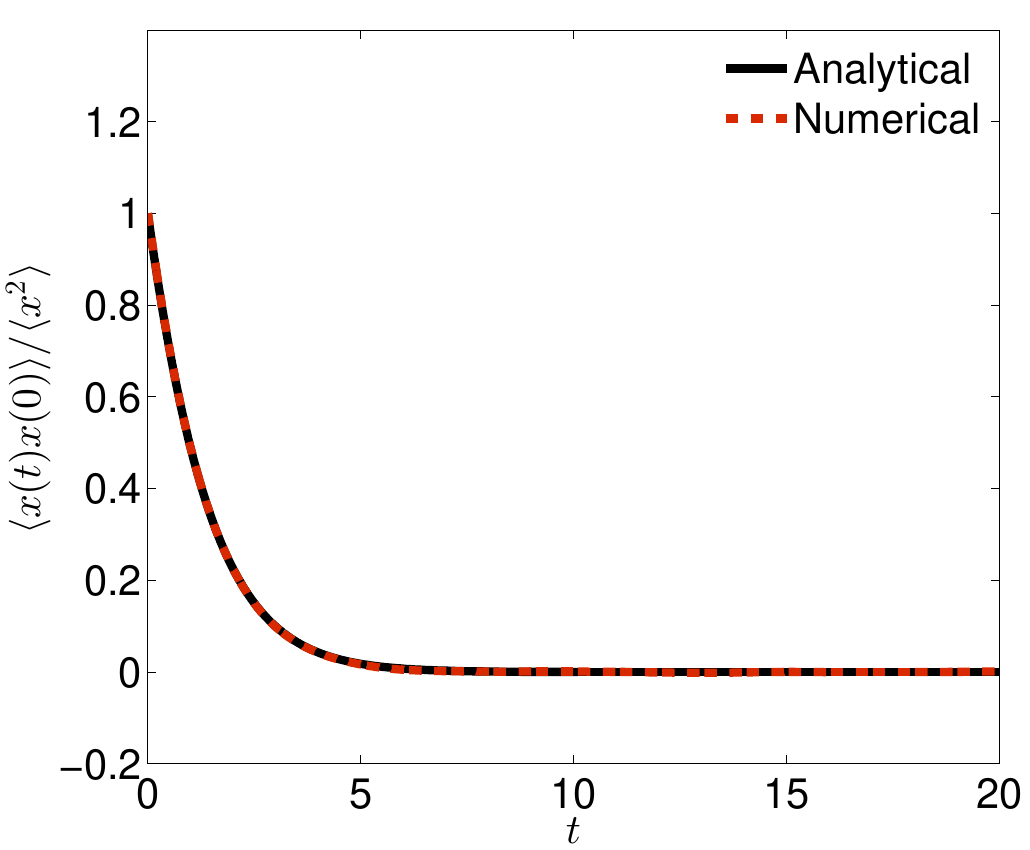}
\caption{\small (Color online) Comparison of various computed (and normalized) time correlation functions of Langevin dynamics without inertia, (i.e., Brownian dynamics), by using the limit method with a stepsize of $h=0.01$ against the analytical solutions derived in~\cref{subsec:Brownian_Nonideal} in solid black lines. The system was simulated for 1000 reduced time units in each case but only the last 80\% of the snapshots were collected to calculate the correlations. Furthermore, 1000 different runs were averaged to reduce the sampling errors. }
\label{fig:Correlations_Brown_Limit_k2_1E3}
\end{figure}

\begin{figure}[tbh]
\centering
\includegraphics[scale=0.4]{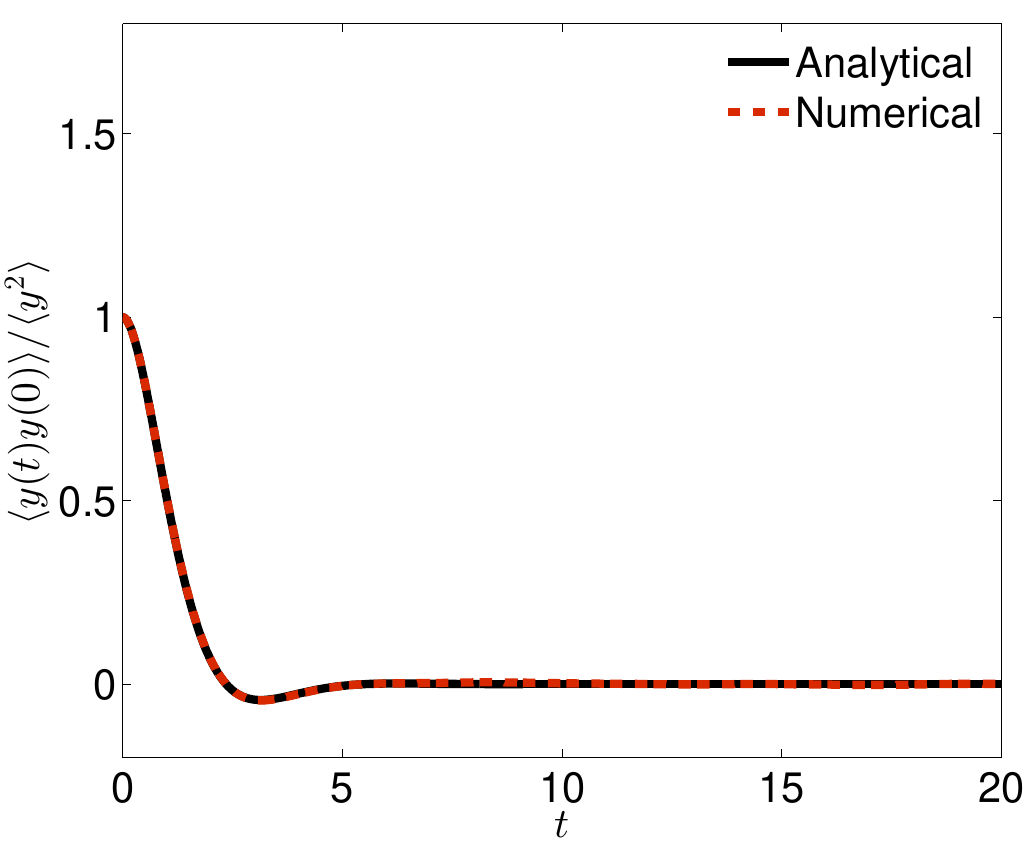}
\includegraphics[scale=0.4]{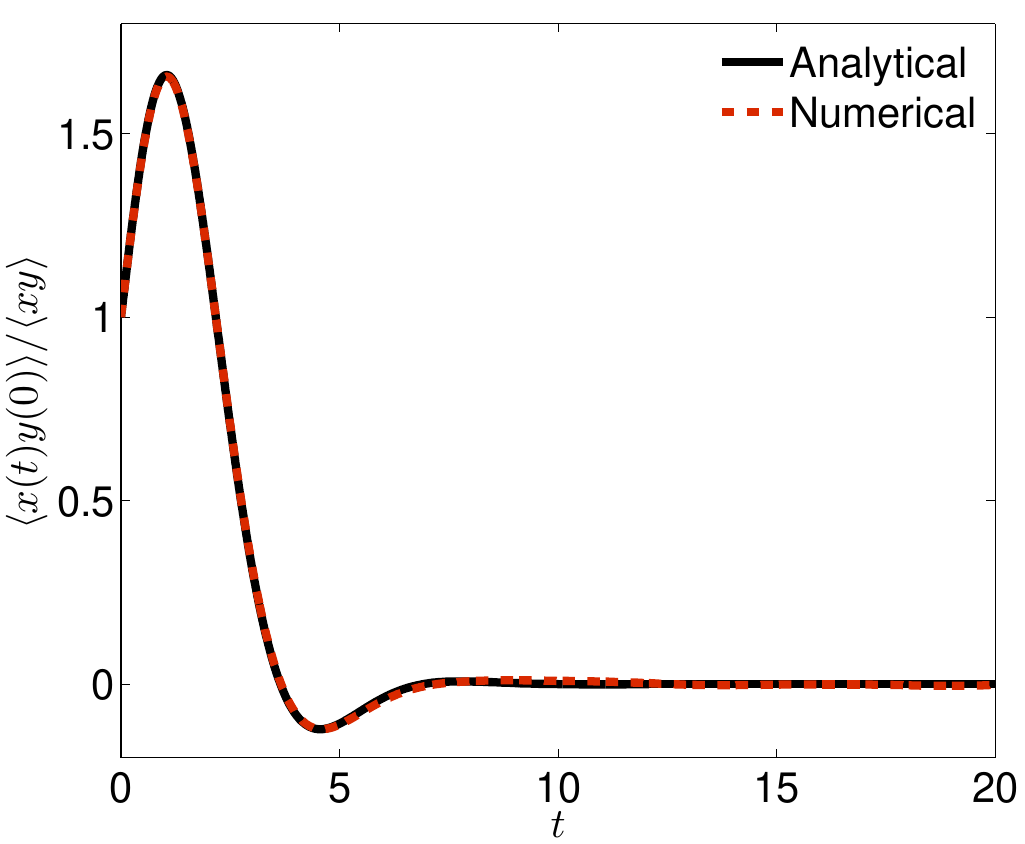}
\includegraphics[scale=0.4]{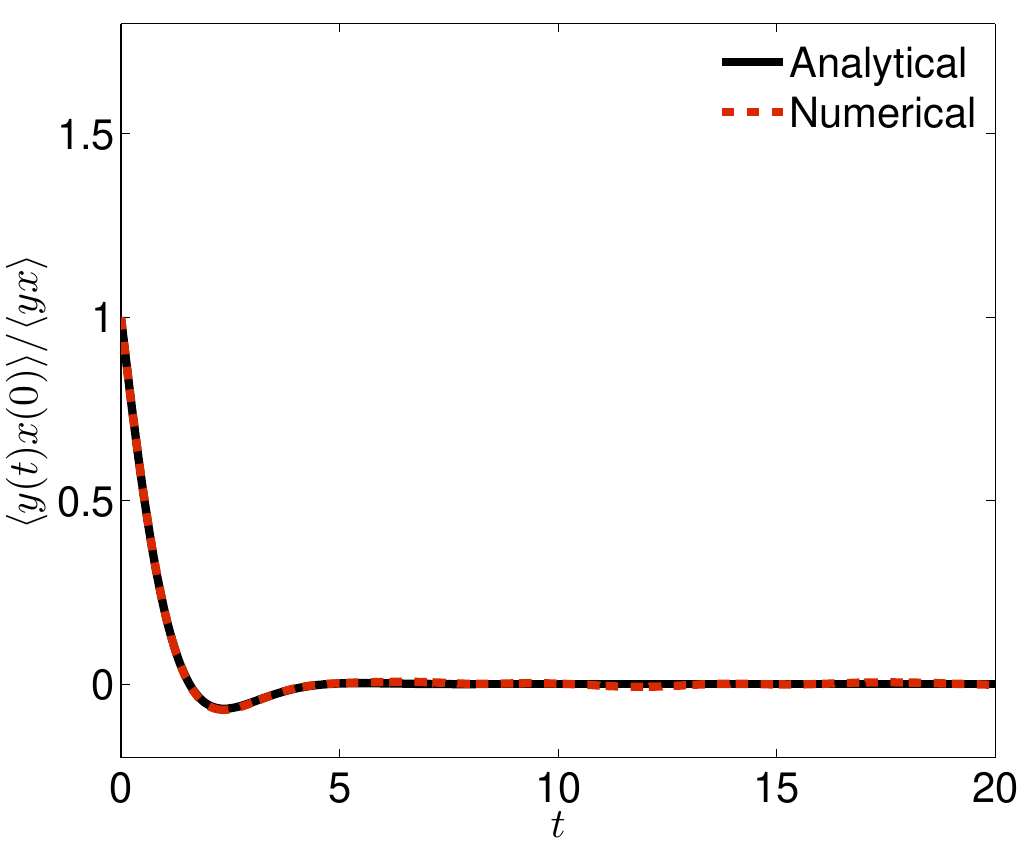}
\includegraphics[scale=0.4]{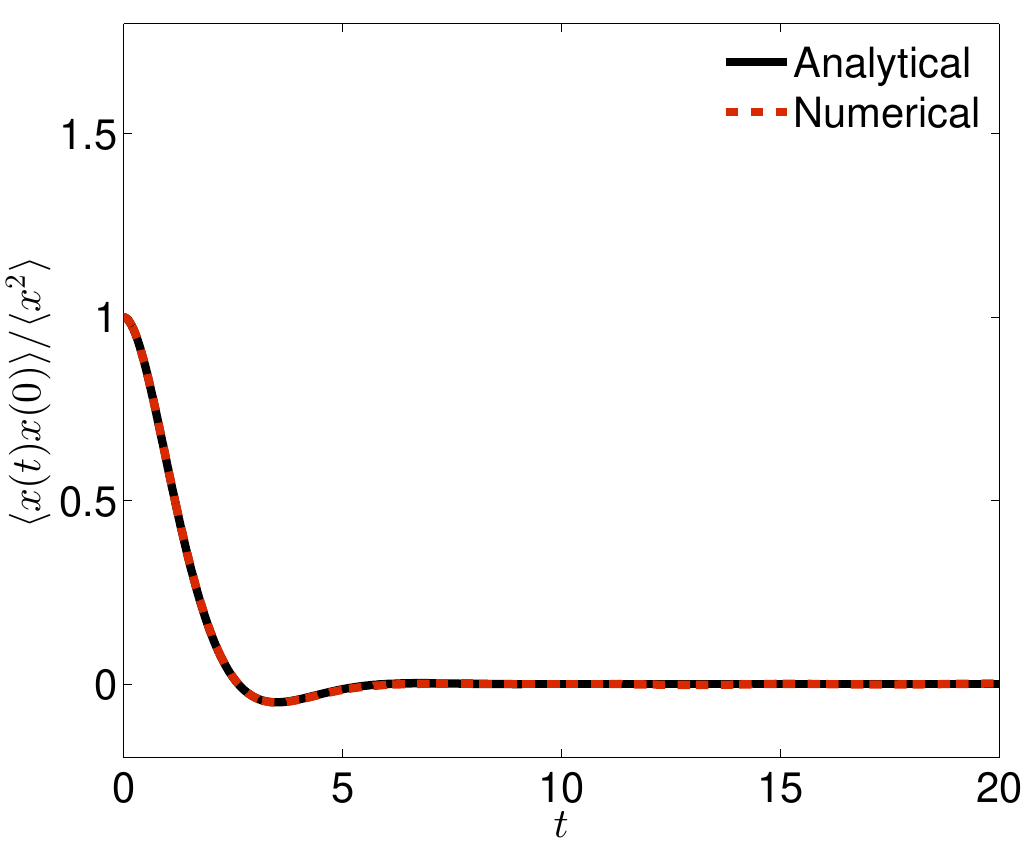}
\caption{\small (Color online) Comparison of various computed (and normalized) time correlation functions of Langevin dynamics with inertia by using the BAOAB method with a stepsize of $h=0.01$ against the analytical solutions derived in~\cref{subsec:Langevin_Nonideal} in solid black lines. The format of the plots is the same as in~\cref{fig:Correlations_Brown_Limit_k2_1E3}. }
\label{fig:Correlations_Langevin_BAOAB_s2_1E3}
\end{figure}

\subsubsection{The BAOAB method}
\label{subsec:BAOAB}

Numerical integration methods, particularly the so-called ``splitting methods'', for Langevin dynamics have been studied systematically in terms of the long term sampling performance by Leimkuhler and coworkers~\cite{Leimkuhler2013,Leimkuhler2013a,Leimkuhler2013c,Leimkuhler2015b,Leimkuhler2014}.  It has been demonstrated that, in terms of sampling configurational quantities, a particular choice of splitting methods, i.e., the ``BAOAB'' method, relying on a Trotter factorization of the stochastic vector field of the original (whole) system into exactly solvable subsystems, is far advantageous to alternative schemes. Subsequently, the optimal design of splitting methods on stochastic dynamics has been studied in a variety of applications~\cite{Leimkuhler2015,Leimkuhler2015a,Leimkuhler2016a,Shang2017}. We point out that the framework of long-time Talay--Tubaro expansion~\cite{Talay1990,Debussche2012,Leimkuhler2013,Leimkuhler2013a,Leimkuhler2013c,Abdulle2014a,Abdulle2014,Leimkuhler2015a,Leimkuhler2015b} can be trivially performed in order to analyse the accuracy of ergodic averages (i.e., averages with respect to the invariant measure) in those systems. We separate the vector field of the Langevin dynamics as
\begin{equation}\label{eq:Splitting_LD}
  \dd \left[ \begin{array}{c} \q \\ \p \end{array} \right] =  \underbrace{\left[ \begin{array}{c} \p \\ {\bf 0} \end{array} \right] \dd t}_\mathrm{A} + \underbrace{\left[ \begin{array}{c} {\bf 0} \\  - s \q \end{array} \right] \dd t }_\mathrm{B} + \underbrace{\left[ \begin{array}{c} {\bf 0} \\ - 2 \left( \p - \mathbf{u} \right) + \bm{\eta} \end{array} \right] }_\mathrm{O} \,,
\end{equation}
where we can solve each piece ``exactly''. That is, both ``A'' and ``B'' pieces can be straightforwardly solved, while it is also possible to derive the exact solution to the Ornstein--Uhlenbeck (``O'') part (solutions in~\cite{Shang2017} for more general settings),
\begin{equation}\label{eq:OU}
  \dd \p = 2 \mathbf{u} \dd t - 2 \p\dd t + \bm{\eta} \,,
\end{equation}
as
\begin{equation}\label{eq:OU_Exact_Sol}
  \p(t) = \mathbf{u} +  \left( \p(0) - \mathbf{u} \right) e^{-2t} + \left( \sqrt{1 - e^{-4t}}/2 \right) \mathrm{\bf R} \,.
\end{equation}
The BAOAB method then can be defined as
\begin{equation}
  e^{h \hat{\mathcal{L}}_\mathrm{BAOAB} } = e^{(h/2) \mathcal{L}_\mathrm{B} } e^{(h/2) \mathcal{L}_\mathrm{A} } e^{h \mathcal{L}_\mathrm{O} } e^{(h/2) \mathcal{L}_\mathrm{A} } e^{(h/2) \mathcal{L}_\mathrm{B} }\,,
\end{equation}
where $\exp\left(h \mathcal{L}_f\right)$ represents the phase space propagator associated with the corresponding vector field $f$. More precisely, the integration steps of the BAOAB method, including the streaming velocity, reads:
\begin{subequations}
\begin{align}
    \p^{n+1/2} &= \p^{n} - h s \q^{n}/2 \,, \\
    \q^{n+1/2} &= \q^{n} + h {\p}^{n+1/2}/2 \,, \\
    \tilde{\p}^{n+1/2} &= \mathbf{u}^{n+1/2} +  \left( \p^{n+1/2} - \mathbf{u}^{n+1/2} \right)e^{-2h} + \left( \sqrt{1 - e^{-4h}}/2 \right) \, \mathbf{R}^{n} \,, \\
    \q^{n+1} &= \q^{n+1/2} + h \tilde{\p}^{n+1/2}/2 \,, \\
    \p^{n+1} &=\tilde{\p}^{n+1/2} - h s \q^{n+1}/2 \,.
\end{align}
\end{subequations}
Note that only one force calculation is required at each step for the BAOAB method (i.e., the force computed at the end of each step will be reused at the start of the subsequent step), which is the same as for alternative schemes, including the SVV method.

\begin{figure}[tbh]
\centering
\includegraphics[scale=0.4]{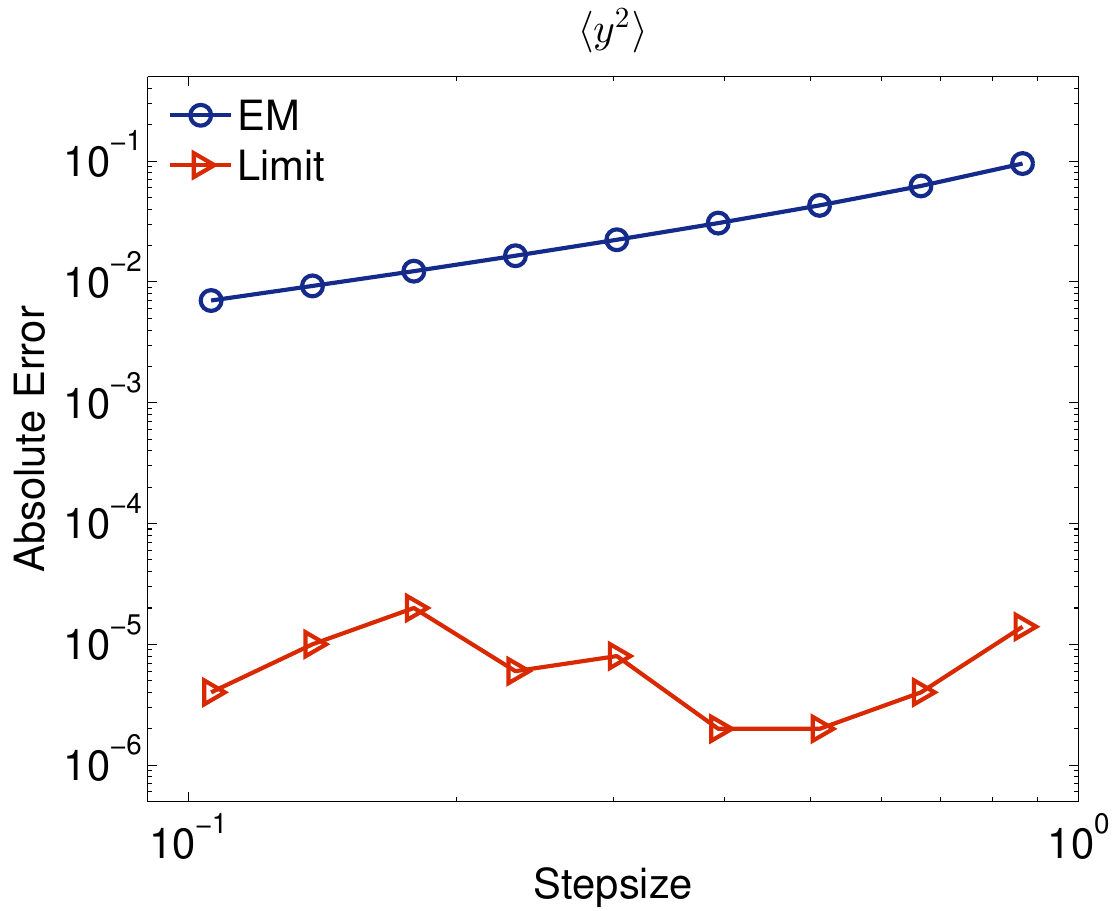}
\includegraphics[scale=0.4]{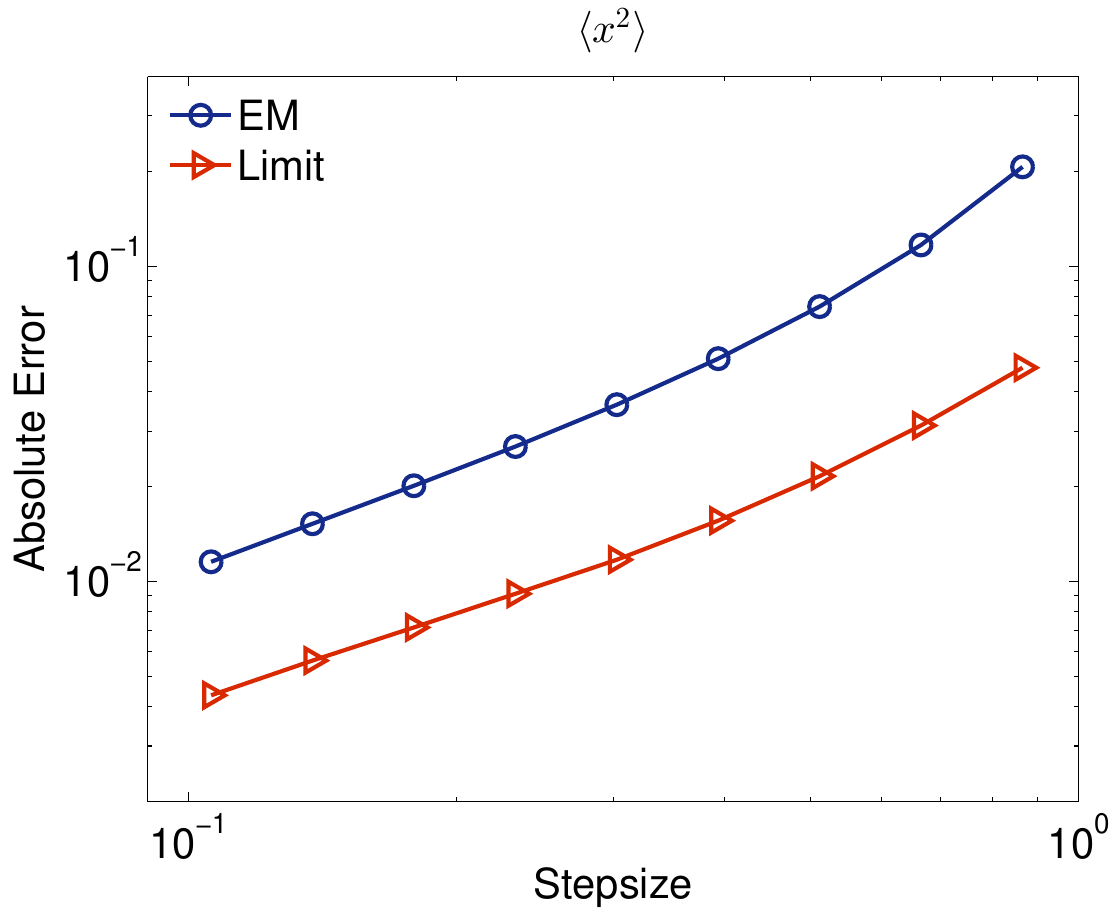}
\caption{\small (Color online) Double logarithmic plot of the computed absolute error in averages $\langle y^2 \rangle$ (left) and $\langle x^2 \rangle$ (right) derived in~\cref{subsec:Brownian_Nonideal} (Brownian dynamics) against stepsize by using the Euler--Maruyama (EM) and limit methods with $\omega=1$ and $D=0.125$. The system was simulated for 1000 reduced time units in each case but only the last 80\% of the snapshots were collected to calculate the static quantities. Furthermore, 100,000 different runs were averaged to reduce the sampling errors. The stepsizes tested began at $h=0.106$ and were increased incrementally by 30\% until substantial errors in correlations were observed. }
\label{fig:Errors_Brown_EM_v_Limit_k2_1E5}
\end{figure}

\begin{figure}[tbh]
\centering
\includegraphics[scale=0.4]{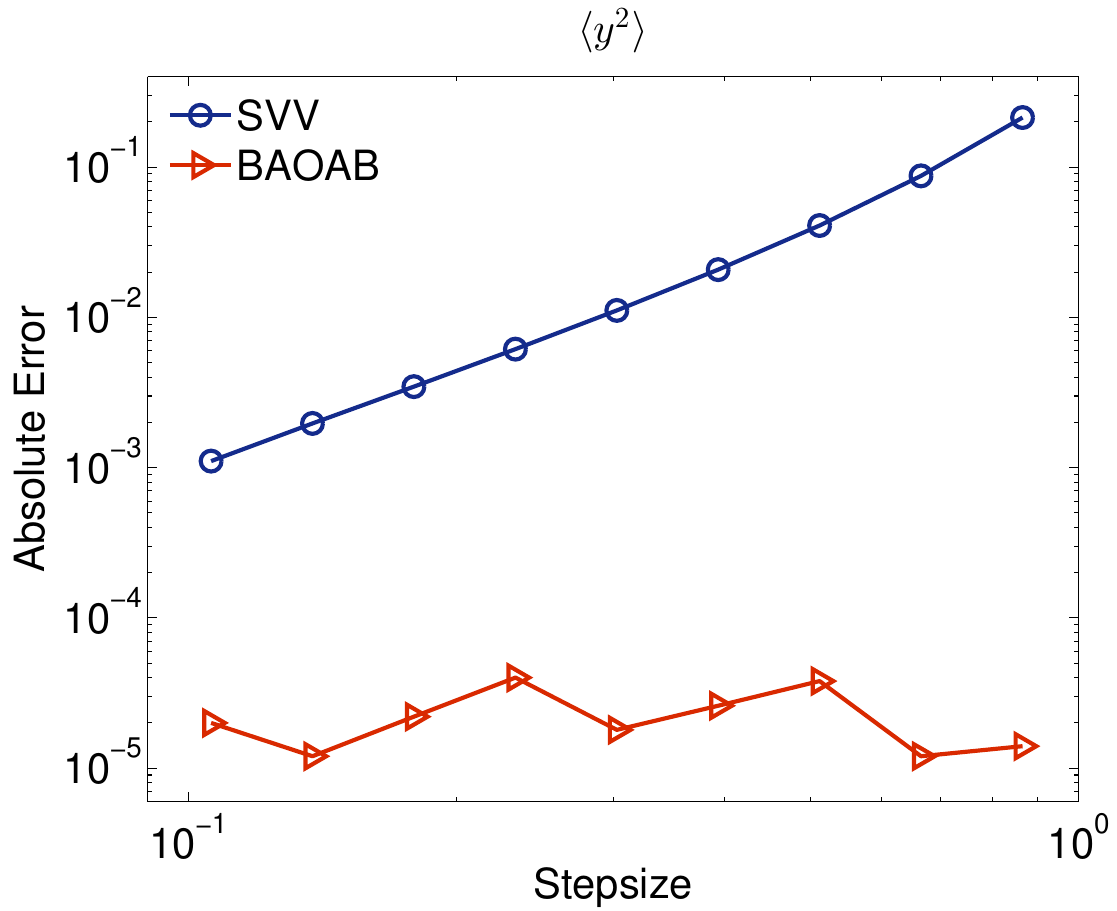}
\includegraphics[scale=0.4]{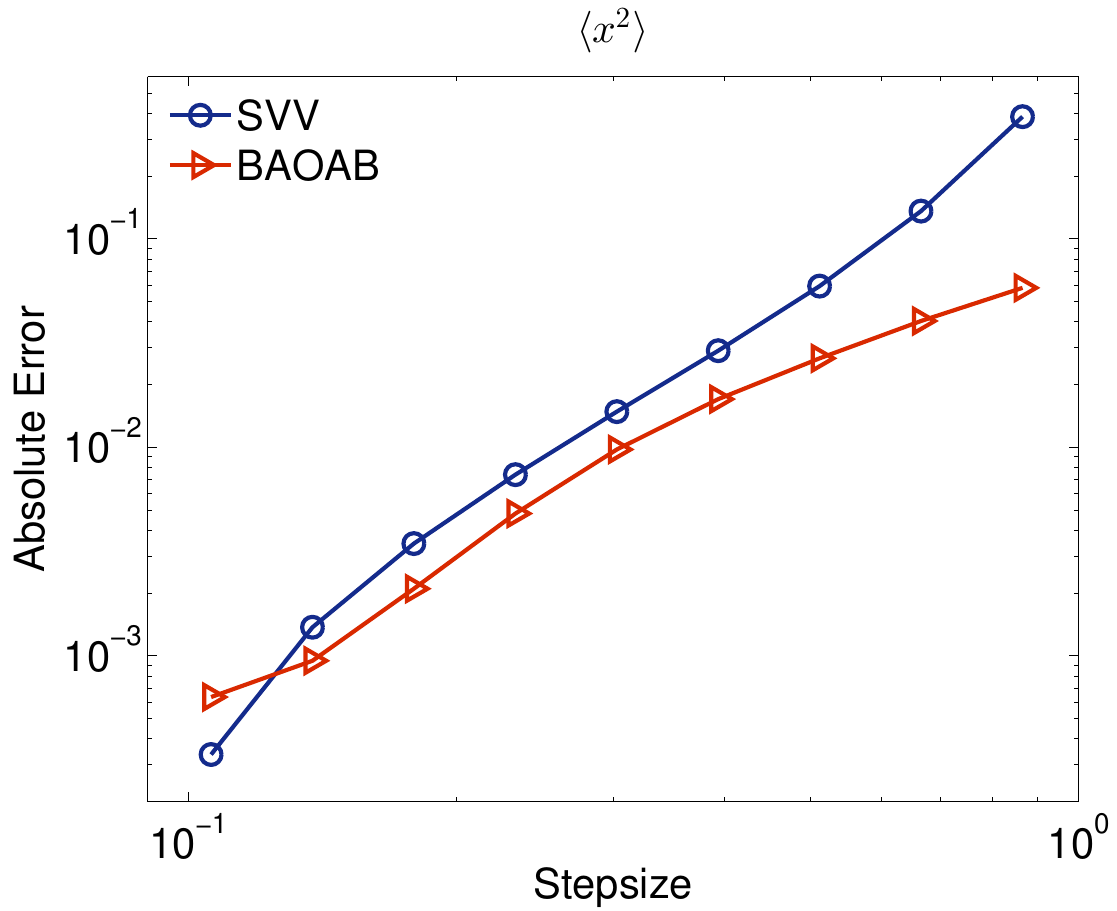}
\caption{\small (Color online) Double logarithmic plot of the  computed absolute error in averages $\langle y^2 \rangle$ (left) and $\langle x^2 \rangle$ (right) derived in~\cref{subsec:Langevin_Nonideal} (Langevin dynamics) against stepsize by using the stochastic velocity Verlet (SVV) and BAOAB methods with $s=2$ and $r=1$. The format of the plots is the same as in~\cref{fig:Errors_Brown_EM_v_Limit_k2_1E5}. }
\label{fig:Errors_Langevin_SVV_v_BAOAB_s2_1E5}
\end{figure}

\section{Numerical experiments}
\label{sec:Numerical_Experiments}

In this section, we conduct a variety of numerical experiments to compare the performance of various methods introduced in~\cref{sec:Numerical_Methods} in noninertia (Brownian) and inertia (Langevin) cases, respectively.

\subsection{Simulation details}
\label{subsec:Simulation_Details}

As described at the beginning of~\cref{sec:Derivations}, we restrict our attention to a single harmonic oscillator of mass $m$ in the presence of a streaming background medium with velocity field $\mathbf{u}$. For the sake of simplicity, we excluded the diagonal contributions from the matrix $\bm{\kappa}$~\cref{eq:kappa} in our numerical experiments. In both cases, the following parameter set was used: $k=2$, $\kB T=0.25$, $\gamma=2$, $\dot{\gamma}=1$, resulting in $\omega=k/\gamma=1$ and $D=\kB T/\gamma=0.125$ in the Brownian case. The mass was set as unity in the Langevin case, thereby leading to $s=2$ and $r=1$. For this choice of parameters the reference $t_\textrm{ref}$ of the Langevin dynamics coincides with the characteristic relaxation time of the inertia-free Brownian case. The initial position of the particle was set at the origin in both cases while the initial momentum in the Langevin case was zero. Unless otherwise stated, the system was simulated for 1000 reduced time units in both cases but only the last 80\% of the data were collected to calculate various quantities derived in~\cref{sec:Derivations}.

\subsection{Results}
\label{subsec:Results}

In order to verify the derivations of the time correlation functions in both noninertia (\cref{subsec:Brownian_Nonideal}) and inertia (\cref{subsec:Langevin_Nonideal}) cases, we plot the computed (and normalized) time correlation functions against the analytical solutions in~\cref{fig:Correlations_Brown_Limit_k2_1E3,fig:Correlations_Langevin_BAOAB_s2_1E3}, respectively. It appears that in both cases the numerical solutions are indistinguishable from the analytical ones with a small stepsize of $h=0.01$. However, as stepsize increases, the time correlation functions do start deviating from the analytical solutions, which leads to the investigation of the accuracy control of average quantities in subsequent figures. We also want to point out that with the same stepsize of $h=0.01$ but a smaller shear rate, say $\dot{\gamma}=0.1$, visible deviations were observed in both cross-correlation functions, i.e., $\left\langle x(t)y(0) \right\rangle / \left\langle xy \right\rangle$ and $\left\langle y(t)x(0) \right\rangle / \left\langle yx  \right\rangle$, while both autocorrelation functions, i.e., $\left\langle y(t)y(0) \right\rangle / \left\langle y^2 \right\rangle$ and $\left\langle x(t)x(0) \right\rangle / \left\langle x^2 \right\rangle$, were still indistinguishable from the analytical solutions. Moreover, the deviations became even stronger if the shear rate was further reduced. This indicates that both cross-correlation functions are more sensitive to the strength of the shear rate.

The accuracy control of average quantities is often used to measure the performance of the numerical methods. To this end, the computed absolute error in averages $\langle y^2 \rangle$ and $\langle x^2 \rangle$ were plotted in~\cref{fig:Errors_Brown_EM_v_Limit_k2_1E5,fig:Errors_Langevin_SVV_v_BAOAB_s2_1E5} for both Brownian and Langevin cases, respectively. (We did not observe significant difference between the methods in both cases in terms of the errors on time correlation functions.) Note that the average $\langle y^2 \rangle$ is actually proportional to the so-called configurational temperature (more discussions in~\cite{Leimkuhler2015,Leimkuhler2016a}), in this case $k\langle y^2 \rangle=\kB T$, which is an important quantity that numerical methods should preserve. The results of $\langle xy \rangle$ were not included due to its sensitivity to sampling errors. To be more specific, in the Brownian case in~\cref{fig:Errors_Brown_EM_v_Limit_k2_1E5}, the limit method is orders of magnitude more accurate than the Euler-Maruyama method in $\langle y^2 \rangle$ while the former still outperforms the latter in $\langle x^2 \rangle$. Although the limit method does not seem to display a second order convergence to the invariant measure as expected in the equilibrium case of $\langle y^2 \rangle$, we point out that it might be very challenging to overcome the impact of sampling errors at such a high level of accuracy with the reference value being $\langle y^2 \rangle=0.125$.

In the case of Langevin dynamics as can be seen in~\cref{fig:Errors_Langevin_SVV_v_BAOAB_s2_1E5}, the BAOAB method is also orders of magnitude more accurate than the stochastic velocity Verlet (SVV) method in $\langle y^2 \rangle$ while the former slightly outperforms the latter in $\langle x^2 \rangle$. Interestingly, in the equilibrium case of $\langle y^2 \rangle$, the accuracy of the BAOAB method does not seem to depend on the stepsize (although it still seems to slightly fluctuate due to the sampling errors at such a high level of accuracy with the reference value again being $\langle y^2 \rangle=0.125$). This behavior is actually consistent with the demonstration in~\cite{Leimkuhler2013a} that the BAOAB method ``exactly'' preserves the average quantity of $\langle y^2 \rangle$ in this particular case.

\section{Summary and Outlook}
\label{sec:Conclusions}

We have derived various time correlation functions and associated quantities of the linear Langevin dynamics (both without and with inertia effects) for a harmonic oscillator in the presence of friction, noise, and an external field with both rotational and deformational contributions.
We have demonstrated how in the nontrivial limit of vanishing mass the inertia results reduce to their noninertia counterparts. While all results were derived explicitly using a most straightforward approach suitable for a classroom, we have mentioned two alternative approaches based on (i) the Fourier transform and (ii) the Fokker--Planck equation. In our numerical experiments, for which algorithms were stated in~\cref{sec:Numerical_Methods}, we not only have verified various time correlation functions~\cref{eq:Langevin_m>0k>0_benchmark_all} derived in this article for the benchmark~\cref{eq:Langevin_Nondim_xy}, but also demonstrated the importance of optimal design of numerical methods. To be more specific, in the Brownian case, we have shown that the limit method substantially outperforms the popular Euler--Maruyama (EM) method in equilibrium while the former appears to be still visibly more accurate than the latter in nonequilibrium. On the other hand in the case of Langevin dynamics, the BAOAB method is orders of magnitude more accurate than the stochastic velocity Verlet (SVV) method in equilibrium whereas the former appears to be only slightly better than the latter in nonequilibrium. While the benchmark~\cref{eq:Langevin_Nondim_xy} involves only dimensionless parameters, we have explicitly stated its connection with dimensional equations from real world applications.
One of them is the study of the full Rouse model~\cite{Rouse1953,Doi1996} (bead-spring chain, i.e., coupled harmonic oscillators with masses, whose eigenmodes behave as harmonic oscillators) for the short-time and high frequency dynamics of unentangled polymeric systems subjected to flows.
With the time correlation functions for ${\bf q}$ obeying~\cref{eq:Langevin} at hand, all relevant properties of a bead-spring chain subjected to flow can be written down upon replacing $m$, $k$, and $\gamma$ by their mode-dependent counterparts $m_p$, $k_p$, and $\gamma_p$~\cite{Doi1996}, where $p=0,1,2,\dots,N$ enumerates the $N$ normal modes of a chain with $N-1$ segments connecting $N$ mass points (beads). In the limit of vanishing mass the known solution of the Rouse model~\cite{Doi1996} is also recovered this way. The analytical methods applied here to solve the linear Langevin dynamics characterized by matrices ${\bf A}$ and ${\bf B}$ in~\cref{matrixA} apply without modification to arbitrary ${\bf A}$ and ${\bf B}$. The numerical methods apply to both linear and nonlinear problems.


\appendix

\section{Nondimensionalization}
\label{app:Nondimensionalization}

In what follows we show that the nondimensionalized version of~\cref{eq:Langevin_xy} is~\cref{eq:Langevin_Nondim_xy}.
Dimensionless quantities $f_{\ast}$ are introduced via $f=f_{\ast} f_\textrm{ref}$, in general, with
reference quantities $f_\textrm{ref}$ carrying the physical dimension. Having restored the asterisks dropped and also rewritten the noise term as a derivative (although it is not rigorously defined in the usual mathematical sense),~\cref{eq:Langevin_Nondim_x} reads
\begin{equation}\label{eq:Langevin_Nondim_x_App}
  \frac{\dd^2 x_{\ast}}{\dd t^2_{\ast}} = -s_x x_{\ast} - 2 \left( \frac{\dd x_{\ast}}{\dd t_{\ast}} - r y_{\ast} \right) + \frac{\dd \mathrm{W}_{{\ast},x}}{\dd t_{\ast}} \,.
\end{equation}
Since $\mathrm{W}^2$ has dimension of time, $\mathrm{W}_\textrm{ref}=\sqrt{t_\textrm{ref}}$,
and~\cref{eq:Langevin_Nondim_x_App}, upon replacing $f_*$ by $f/f_\textrm{ref}$,
and subsequent multiplication by
by $mq_\textrm{ref}/t_\textrm{ref}^2$ on both sides of the equation yields
\begin{equation}\label{eq:Langevin_Nondim_x_App_2}
  m \frac{\dd^2x}{\dd t^2}
  = -s_x \frac{m x}{t_\textrm{ref}^2} - 2 \left( \frac{m}{t_\textrm{ref}}\frac{\dd x}{\dd t}
  - r \frac{my}{t_\textrm{ref}^2} \right) + m q_\textrm{ref} \frac{t_\textrm{ref}^{1/2}}{t_\textrm{ref}^2} \frac{\dd\mathrm{W}_x}{\dd t} \,.
\end{equation}
\begin{proof}Inserting $q_\textrm{ref}$, $t_\textrm{ref}$, $s_\mu$, and $r$  from~\cref{eq:reference_quantities} and~\cref{eq:sr}
into~\cref{eq:Langevin_Nondim_x_App_2}
\begin{align}
  m \frac{\dd^2x}{\dd t^2}
  &= - \frac{4mk_x}{\gamma^2} \frac{m \gamma^2 x}{4 m^2} - 2 \left( \frac{m\gamma}{2m}\frac{\dd x}{\dd t}
  - \frac{2m\dot{\gamma}}{\gamma} \frac{m\gamma^2 y}{4m^2} \right) + m \frac{2^{3/2}\sigma\sqrt{m}}{\gamma^{3/2}}  \left(\frac{\gamma}{2m}\right)^{3/2} \frac{\dd\mathrm{W}_x}{\dd t}
  \nonumber \\
  &= - k_x x - \gamma\left( \frac{\dd x}{\dd t}
  -  \dot{\gamma} y \right) + \sigma \frac{\dd\mathrm{W}_x}{\dd t}\,.
\end{align}\end{proof}

\section{Ideal Brownian dynamics: $m=0, k_x=k_y=0$}

\subsection{Time correlation function $\left\langle [x(t)-x(0)][y(t)-y(0)] \right\rangle$}
\label{app:moko_xy}

\begin{proof}Starting from~\cref{eq:Langevin_m0k0_xy}, with the help of~\cref{eq:Wiener}, we arrive at~\cref{3.5} as follows
\begin{align}
  \left\langle [x(t)\!-\!x(0)][y(t)\!-\!y(0)] \right\rangle &= \left\langle \int_0^t \dot{x}(t_1) \, \dd t_1 \int_0^t \dot{y}(t_2) \, \dd t_2 \right\rangle
  \nonumber \\
  &= \sqrt{2D} \int_0^t \int_0^t \left[ \dot{\gamma} \left\langle y(t_1) \eta_y(t_2) \right\rangle\!+\!\sqrt{2D}\, \left\langle \eta_x(t_1) \, \eta_y(t_2) \right\rangle \right] \dd t_2\dd t_1
  \nonumber \\
  &= 2D \dot{\gamma} \int_0^t \int_0^t \int_0^{t_1} \left\langle \eta_y(t'_1) \eta_y(t_2) \right\rangle \dd t'_1 \, \dd t_2 \, \dd t_1
  \nonumber \\
  &= 2D \dot{\gamma} \int_0^t \int_0^t \int_0^{t_1} \delta(t'_1-t_2) \, \dd t'_1 \, \dd t_2 \, \dd t_1
  \nonumber \\
  &= 2D \dot{\gamma} \int_0^t \int_0^{t_1} \int_0^{t_1} \delta(t'_1-t_2) \, \dd t'_1 \, \dd t_2 \, \dd t_1
  \nonumber \\
  &= 2D \dot{\gamma} \int_0^t \int_0^{t_1} \, \dd t_2 \, \dd t_1 = D \dot{\gamma} t^2 \,.
\end{align}
\end{proof}

\subsection{Mean squared displacement $\left\langle [x(t)-x(0)]^2 \right\rangle$}
\label{app:moko_xx}

\begin{proof}Starting from~\cref{eq:Langevin_m0k0_xy}, with the help of~\cref{eq:Wiener}, we arrive at~\cref{eq:Langevin_m0k0_MSD_xx} as follows
\begin{align}
  \left\langle \left[x(t)-x(0)\right]^2 \right\rangle &= \left\langle \int_0^t \dot{x}(t_1) \, \dd t_1 \int_0^t \dot{x}(t_2) \, \dd t_2 \right\rangle
  \nonumber \\
  &= \int_0^t \int_0^t \left[ 2D \, \delta(t_1-t_2) + \dot{\gamma}^2 \left\langle y(t_1)y(t_2) \right\rangle \right] \dd t_1 \, \dd t_2
  \nonumber \\
  &= 2Dt + \dot{\gamma}^2 \int_0^t \int_0^t \left[ 2D\textrm{min}(t_1,t_2) + y^2_0 \right] \dd t_1 \, \dd t_2
  \nonumber \\
  &= 2Dt + 2D \dot{\gamma}^2 \left( \int_0^t \int_{t_2}^t t_2 \dd t_1 \dd t_2 + \int_{0}^t \int_0^{t_2} \!t_1 \dd t_1 \dd t_2 \right) + \left( \dot{\gamma} y_0 t \right)^2
  \nonumber \\
  &= 2D t \left[ 1 + \frac{1}{3} \left( \dot{\gamma} t \right)^2 \right] + \left( \dot{\gamma} y_0 t \right)^2 \,.
\end{align}
\end{proof}

\section{Nonideal Brownian dynamics: $m=0, k_x,k_y>0$}

\subsection{Time correlation function $\left\langle y(t_1)y(t_2) \right\rangle$}
\label{app:mok>o_yt1yt2}

\begin{proof}Starting from~\cref{eq:Langevin_m0k>0_sol_y}, with the help of~\cref{eq:Wiener}
together with the identity $\min(t_1,t_2) = (t_1+t_2)/2 - |t_1-t_2|/2$,
\cref{eq:Langevin_m0k>0_yt1yt2} is obtained as follows
\begin{align}
  \left\langle y(t_1)y(t_2) \right\rangle &= 2D \left\langle \int_{-\infty}^{t_1} \, \eta_y(t'_1) e^{-\omega_y (t_1-t'_1)} \, \dd t'_1 \int_{-\infty}^{t_2} \, \eta_y(t'_2) e^{-\omega_y (t_2-t'_2)} \, \dd t'_2 \right\rangle
  \nonumber \\
  &= 2D \int_{-\infty}^{t_1} \int_{-\infty}^{t_2}  e^{-\omega_y(t_1+t_2-t'_1-t'_2)} \left\langle \eta_y(t'_1) \, \eta_y(t'_2) \right\rangle \dd t'_2 \, \dd t'_1
  \nonumber \\
  &= 2D \int_{-\infty}^{t_1} \int_{-\infty}^{t_2}  e^{-\omega_y(t_1+t_2-t'_1-t'_2)} \, \delta(t'_1-t'_2) \, \dd t'_2 \, \dd t'_1
  \nonumber \\
  &= 2D \int_{-\infty}^{\min(t_1,t_2)} e^{-\omega_y(t_1+t_2-2t'_1)} \, \dd t'_1
  \nonumber \\
  &= \frac{D}{\omega_y} e^{-\omega_y |t_1-t_2|} \,,
\end{align}
\end{proof}

\subsection{Time correlation function $\left\langle x(t)y(0) \right\rangle$}
\label{app:mok>o_xty0}

\begin{proof}Starting from~\cref{eq:Langevin_m0k>0_sol_y}, an intermediate result is
\begin{align}
  \left\langle y(t_1) \, \eta_y(t_2) \right\rangle &= \sqrt{2D}\int_{-\infty}^{t_1} \left\langle \eta_y(t'_1) \, \eta_y(t_2) \right\rangle e^{-\omega_y (t_1-t'_1)} \, \dd t'_1 \nonumber \\
  &= \sqrt{2D}\int_{-\infty}^{t_1} \, \delta(t'_1-t_2) e^{-\omega_y(t_1-t'_1)} \, \dd t'_1
  \nonumber \\
  &= \sqrt{2D}\,e^{-\omega_y (t_1-t_2)}\Theta(t_1-t_2) \,,
  \label{C.3}
\end{align}
where $\Theta$ denotes the Heaviside step function. Since $\left\langle \eta_x(t) \, \eta_y(t') \right\rangle = 0$, one recovers~\cref{eq:Langevin_m0k>0_xty0} using~\cref{C.3}
\begin{align}
  \left\langle x(t)y(0) \right\rangle &= \sqrt{2D}\int_{-\infty}^t  \int_{-\infty}^0 \left\langle \left[\dot{\gamma}y(t_1)+ \sqrt{2D} \,\eta_x(t_1)\right] \eta_y(t_2)\right\rangle  e^{-\omega_x (t-t_1)+\omega_y t_2} \, \dd t_2 \, \dd t_1 \nonumber \\
  &= 2D\dot{\gamma}\int_{-\infty}^t \int_{-\infty}^0 e^{-\omega_y (t_1-t_2)} \Theta(t_1-t_2) e^{-\omega_x (t-t_1)+\omega_y t_2} \, \dd t_2 \, \dd t_1
  \nonumber \\
  &= 2D\dot{\gamma} e^{-\omega_x t}  \int_{-\infty}^0 e^{2\omega_y t_2} \int_{t_2}^t e^{(\omega_x-\omega_y) t_1} \, \dd t_1 \, \dd t_2\,
  \nonumber \\
  &= D\dot{\gamma}\,
  \frac{(\omega_x+\omega_y)e^{-\omega_y t}-2\omega_y e^{-\omega_x t}}{(\omega_x^2-\omega_y^2)\omega_y} \,.
\end{align}
\end{proof}

\subsection{Time correlation function $\left\langle y(t)x(0) \right\rangle$}
\label{app:mok>o_ytx0}

\begin{proof}In full analogy to~\cref{app:mok>o_xty0},~\cref{eq:Langevin_m0k>0_ytx0} is derived via
\begin{align}
  \left\langle y(t)x(0) \right\rangle &= \sqrt{2D}\int_{-\infty}^t  \int_{-\infty}^0 \!\left\langle \eta_y(t_1) \left[\dot{\gamma}y(t_2)+ \sqrt{2D} \, \eta_x(t_2)\right]\right\rangle  e^{-\omega_y (t-t_1)+\omega_x t_2} \, \dd t_2 \, \dd t_1 \nonumber \\
  &= 2D\dot{\gamma}\int_{-\infty}^t \int_{-\infty}^0 e^{-\omega_y (t_2-t_1)} \Theta(t_2-t_1) e^{-\omega_y (t-t_1)+\omega_x t_2} \, \dd t_2 \, \dd t_1
  \nonumber \\
  &= 2D\dot{\gamma} e^{-\omega_y t} \int_{-\infty}^0 e^{(\omega_x-\omega_y) t_2} \int_{-\infty}^{t_2} e^{2\omega_y t_1} \, \dd t_1 \, \dd t_2\,
  \nonumber \\
  &= \frac{D\dot{\gamma}e^{-\omega_y t}}{(\omega_x+\omega_y)\omega_y} \,.
\end{align}
\end{proof}

\subsection{Time correlation function $\left\langle x(t)x(0) \right\rangle$}
\label{app:mok>o_xtx0}

\begin{proof}The solution~\cref{eq:Langevin_m0k>0_sol_x} can be written as the sum of two uncorrelated contributions $x(t) = x_1(t) + x_2(t)$, where $x_1(t)$ and $x_2(t)$ are given by
\begin{equation}\label{eq:Langevin_m0k>0_sol_x1}
  x_1(t) = \dot{\gamma} \int_{-\infty}^t y(t') e^{-\omega_x (t-t')} \, \dd t' \,, \quad x_2(t) = \sqrt{2D} \int_{-\infty}^t \, \eta_x(t') e^{-\omega_x (t-t')} \, \dd t' \,.
\end{equation}
While $\left\langle x_2(t)x_2(0) \right\rangle$ can be immediately obtained from~\cref{eq:Langevin_m0k>0_sol_y} and~\cref{eq:Langevin_m0k>0_yty0} as
\begin{equation}\label{eq:Langevin_m0k>0_xt2x20}
  \left\langle x_2(t)x_2(0) \right\rangle = \frac{D e^{-\omega_x t}}{\omega_x} \,,
\end{equation}
and since the cross-correlation $\langle x_1(t)x_2(0)\rangle$ vanishes for all $t$ as $\langle \eta_x(t)\eta_y(0)\rangle$ does,
the remaining contribution to $\langle x(t)x(0)\rangle$ is
\begin{align}
  \left\langle x_1(t)x_1(0) \right\rangle
  &= \dot{\gamma}^2 \int_{-\infty}^t \int_{-\infty}^0 \left\langle y(t_1) y(t_2) \right\rangle e^{-\omega_x (t-t_1-t_2)} \, \dd t_2 \, \dd t_1
  \nonumber \\
  &= \frac{\dot{\gamma}^2 D}{\omega_y} \int_{-\infty}^t  \int_{-\infty}^0 e^{-\omega_y |t_1-t_2|} e^{-\omega_x (t-t_1-t_2)} \, \dd t_2 \, \dd t_1
  \nonumber \\
  &= \frac{\dot{\gamma}^2 D e^{-\omega_x t}}{\omega_y} \left[ \int_{-\infty}^0 e^{(\omega_x+\omega_y) t_2} \int_{t_2}^t e^{(\omega_x-\omega_y) t_1} \, \dd t_1 \, \dd t_2 \right.\nonumber \\
  & \quad\qquad\qquad \left. + \int_{-\infty}^0 e^{(\omega_x-\omega_y) t_2} \int_{-\infty}^{t_2} e^{(\omega_x+\omega_y) t_1} \, \dd t_1 \, \dd t_2 \right]
  \nonumber \\
  &= \frac{\dot{\gamma}^2 D \left( \omega_x e^{-\omega_y t}-\omega_y e^{-\omega_x t} \right) }{(\omega_x^2-\omega_y^2)\omega_x\omega_y} \,.
  \label{eq:Langevin_m0k>0_xt1x10}
\end{align}
The sum of~\cref{eq:Langevin_m0k>0_xt2x20} and~\cref{eq:Langevin_m0k>0_xt1x10} is
the desired expression~\cref{eq:Langevin_m0k>0_xtx0} for $\langle x(t)x(0)\rangle$.
\end{proof}

\section{Ideal Langevin dynamics: $m>0, k_x=k_y=0$}

\subsection{Mean squared displacement $\left\langle [y(t)-y(0)]^2 \right\rangle$}
\label{app:m>oko_yy}

\begin{proof}Rewriting $y(t)-y(0)$ as an integral, using~\cref{eq:Langevin_m>0k0_sol_y}, we arrive at~\cref{eq:Langevin_m>0k0_MSD} as follows
\begin{align}
  \left\langle [y(t)-y(0)]^2 \right\rangle &= \left\langle \int_0^t \dot{y}(t_1) \, \dd t_1 \int_0^t \dot{y}(t_2) \, \dd t_2 \right\rangle
  \nonumber \\
  &= \int_0^t \int_0^t \left\langle \dot{y}(t_1) \dot{y}(t_2) \right\rangle \dd t_1 \, \dd t_2
  = \frac{1}{4} \int_0^t \int_0^t e^{-2|t_1-t_2|} \, \dd t_1 \, \dd t_2
  \nonumber \\
  &= \frac{1}{4} \left[ \int_0^t e^{-2t_1} \int_0^{t_1} e^{2t_2} \, \dd t_2 \, \dd t_1  + \int_0^t e^{2t_1} \int_{t_1}^t e^{-2t_2} \, \dd t_2 \, \dd t_1 \right]
  \nonumber \\
  &
  = \frac{1}{8} \left( 2t + e^{-2t} - 1 \right) \,.
\end{align}
\end{proof}

\section{Nonideal Langevin dynamics: $m>0, k\equiv k_x=k_y>0$}

\subsection{Solution of the system $y(t)$}
\label{app:m>ok>o_y}

\begin{proof}Let
\begin{equation}
   G^{\pm}_y(t) = G_y \left(t,s_\pm\right) = e^{-s_\pm t} \int_{-\infty}^t e^{s_\pm t'} \eta_y(t') \, \dd t' \,,
\end{equation}
Equation~\cref{eq:Langevin_m>0k>0_sol_y} may be rewritten as
\begin{equation}
  2\sqrt{1-s} y = G^{-}_y - G^{+}_y \,.
\end{equation}
Differentiating this expression with respect $t$ gives
\begin{equation}
  2\sqrt{1-s}\, \dot{y} = - s_- G^{-}_y + s_+ G^{+}_y \,,
\end{equation}
and differentiating once more with respect to $t$ gives
\begin{equation}
  2\sqrt{1-s} \,\ddot{y} = s^2_- G^{-}_y - s^2_+ G^{+}_y + (s_+ - s_-) \eta_y \,.
\end{equation}
Substituting the above three equations into~\cref{eq:Langevin_m>0k>0_y} we have proven~\cref{eq:Langevin_m>0k>0_sol_y}.
\end{proof}

\subsection{Time correlation function $\left\langle y(t)y(0) \right\rangle$}
\label{app:m>ok>o_yty0}

\begin{proof}
We begin with the intermediate result
\begin{align}
  \left\langle G_y(t,a) G_y(0,b) \right\rangle
  &= \left\langle\int_{-\infty}^t e^{-a(t-t_1)} \, \eta_y(t_1) \, \dd t_1
  \int_{-\infty}^0 e^{-b(0-t_2)} \, \eta_y(t_2) \, \dd t_2 \right\rangle
  \nonumber \\
  &= \int_{-\infty}^t \int_{-\infty}^0 e^{-a(t-t_1)+bt_2}
  \left\langle \eta_y(t_1) \, \eta_y(t_2) \right\rangle \dd t_2 \, \dd t_1
  \nonumber \\
  &= \int_{-\infty}^t \int_{-\infty}^0 e^{-a(t-t_1)+bt_2} \, \delta(t_1-t_2) \, \dd t_2 \, \dd t_1
  \nonumber \\
  &= e^{-at} \int_{-\infty}^0 e^{(a+b)t_1} \, \dd t_1
  \nonumber \\
  &= \frac{e^{-at}}{a+b} \,, \quad \left[ \Re(a+b)>0 \right].
  \label{E.5}
\end{align}
Given $a,b \in \{s_-, s_+\}$, one can verify that the real parts of $a+b$ are always positive. Therefore,
starting from~\cref{eq:Langevin_m>0k>0_sol_y} one approves~\cref{eq:Langevin_m>0k>0_benchmark_yy} with the help of~\cref{E.5}
\begin{align}\label{eq35}
  \left\langle y(t)y(0) \right\rangle
  &= \frac{1}{4(1-s)} \left\langle \left[ G_y(t,s_-) - G_y(t,s_+) \right] \left[ G_y(0,s_-) - G_y(0,s_+) \right] \right\rangle
  \nonumber \\
  &= \frac{1}{4(1-s)} \left[ \frac{e^{-s_- t}}{2s_-} - \frac{e^{-s_- t}}{s_-+s_+} - \frac{e^{-s_+ t}}{s_-+s_+} + \frac{e^{-s_+ t}}{2s_+} \right]
  \nonumber \\
  &= \frac{1}{8 s\sqrt{1-s}} (C_1^+ + C_1^-) \,.
\end{align}
\end{proof}

\subsection{Time correlation function $\left\langle x(t)y(0) \right\rangle$}
\label{app:m>ok>o_xty0}

\begin{proof}We need the following intermediate results,
\begin{align}
  \left\langle G_y(t_1,s') \, \eta_y(t_2) \right\rangle
  &= \left\langle \int_{-\infty}^{t_1} e^{-s'(t_1-t')} \, \eta_y(t') \, \eta_y(t_2) \, \dd t' \right\rangle
  \nonumber \\
  &= \int_{-\infty}^{t_1} e^{-s'(t_1-t')}
  \left\langle \eta_y(t') \, \eta_y(t_2) \right\rangle \dd t'
  \nonumber \\
  &= \int_{-\infty}^{t_1} e^{-s'(t_1-t')} \, \delta(t'-t_2) \, \dd t'
  \nonumber \\
  &= e^{-s'(t_1-t_2)} \Theta(t_1-t_2) \,,
  \label{E.7}
\end{align}
and, with $y(t)$ from~\cref{eq:Langevin_m>0k>0_sol_y},
\begin{align}
  \left\langle y(t_1) \, \eta_y(t_2) \right\rangle
  &= \frac{1}{2\sqrt{1-s}} \left\langle \left[ G_y(t_1,s_-) - G_y(t_1,s_+) \right] \eta_y(t_2) \right\rangle
  \nonumber \\
  &= \frac{1}{2\sqrt{1-s}} \left[ \left\langle G_y(t_1,s_-) \, \eta_y(t_2) \right\rangle - \left\langle G_y(t_1,s_+) \, \eta_y(t_2) \right\rangle \right]
  \nonumber \\
  &= \frac{1}{2\sqrt{1-s}} \left[ e^{-s_-(t_1-t_2)} - e^{-s_+(t_1-t_2)} \right] \Theta(t_1-t_2) \,.
  \label{E.8}
\end{align}
Using~\cref{E.7} and~\cref{E.8}, with yet unspecified $a$ and $b$
\begin{align}
  \left\langle G_x(t,a) G_y(0,b) \right\rangle
  &= \left\langle \int_{-\infty}^t e^{-a(t-t_1)} \left[ 2ry(t_1) + \eta_x(t_1) \right] \dd t_1
  \int_{-\infty}^0 e^{-b(0-t_2)} \, \eta_y(t_2) \, \dd t_2 \right\rangle
  \nonumber \\
  &= 2r \int_{-\infty}^t  \int_{-\infty}^0 e^{-a(t-t_1)+bt_2}
  \left\langle y(t_1) \, \eta_y(t_2) \right\rangle \dd t_2 \, \dd t_1
  \nonumber \\
  &= \frac{r}{\sqrt{1-s}} \int_{-\infty}^t \int_{-\infty}^0  e^{-a(t-t_1)+bt_2} \left[ e^{-s_-(t_1-t_2)} - e^{-s_+(t_1-t_2)} \right] \Theta(t_1\!-\!t_2)\dd t_2\dd t_1
  \nonumber \\
  &= \frac{re^{-at}}{\sqrt{1-s}} \int_{t_2}^t \int_{-\infty}^0  e^{at_1+bt_2} \left[ e^{-s_-(t_1-t_2)} - e^{-s_+(t_1-t_2)} \right] \dd t_2 \, \dd t_1
  \,.
  \label{E.9}
\end{align}
For all the relevant choices of $a$ and $b$ in~\cref{E.9}, the integrals can be performed
\begin{align*}
  \left\langle G_x(t,s_-) G_y(0,s_-) \right\rangle
  &= \frac{re^{-s_-t}}{\sqrt{1-s}} \!\left[ \frac{t}{2s_-} + \frac{1}{4s^2_-} - \frac{1}{s_--s_+} \left( \frac{e^{(s_--s_+)t}}{s_-+s_+} - \frac{1}{2s_-} \right) \right], \nonumber \\
  \left\langle G_x(t,s_-) G_y(0,s_+) \right\rangle &= \frac{re^{-s_-t}}{\sqrt{1-s}}\! \left[ \frac{t}{s_-+s_+} + \frac{1}{(s_-+s_+)^2} - \frac{1}{s_--s_+} \left( \frac{e^{(s_--s_+)t}}{2s_+} - \frac{1}{s_-+s_+} \right) \right], \nonumber \\
  \left\langle G_x(t,s_+) G_y(0,s_-) \right\rangle &= \frac{re^{-s_+t}}{\sqrt{1-s}}\! \left[ \frac{1}{s_+-s_-} \!\left( \frac{e^{(s_+-s_-)t}}{2s_-} - \frac{1}{s_-+s_+} \right) - \frac{t}{s_-+s_+} - \frac{1}{(s_-+s_+)^2} \right],
  \nonumber \\
  \left\langle G_x(t,s_+) G_y(0,s_+) \right\rangle &= \frac{re^{-s_+t}}{\sqrt{1-s}}\! \left[ \frac{1}{s_+-s_-} \! \left( \frac{e^{(s_+-s_-)t}}{s_-+s_+} - \frac{1}{2s_+} \right) - \frac{t}{2s_+} - \frac{1}{4s^2_+} \right],
\end{align*}
With their help the correlation $\langle x(t)y(0)\rangle$ can now be calculated quite conveniently as
\begin{align}
  \left\langle x(t)y(0) \right\rangle
  &= \frac{1}{4(1-s)} \left\langle \left[ G_x(t,s_-) - G_x(t,s_+) \right] \left[ G_y(0,s_-) - G_y(0,s_+) \right] \right\rangle  \nonumber \\
  &= \frac{re^{-s_+t}}{8(1-s)^{3/2}} \frac{\sqrt{1-s}}{\left(1+\sqrt{1-s}\right)} \left[ t + \frac{1}{2}\left( 1 + \frac{1}{1+\sqrt{1-s}} \right) + \frac{1}{\sqrt{1-s}}  \right]
  \nonumber \\
  & \quad - \frac{re^{-s_-t}}{8(1-s)^{3/2}} \frac{\sqrt{1-s}}{\left(1-\sqrt{1-s}\right)} \left[ - t - \frac{1}{2}\left( 1 + \frac{1}{1-\sqrt{1-s}}  \right) + \frac{1}{\sqrt{1-s}} \right].
\end{align}
Multiplying $\left(1-\sqrt{1-s}\right)^2\left(1+\sqrt{1-s}\right)^2=s^2$ on both sides gives
\begin{align}
  s^2 \left\langle x(t)y(0) \right\rangle
  &= \frac{re^{-s_+t}}{8(1-s)^{3/2}} \left(1-\sqrt{1-s}\right)^2 \left[ \left(2+t\right)\sqrt{1-s} + \left(\frac{1}{2}+t\right) \left(1-s\right) + 1  \right] \nonumber \\
  & \quad - \frac{re^{-s_-t}}{8(1-s)^{3/2}} \left(1+\sqrt{1-s}\right)^2 \left[ - \left(2+t\right)\sqrt{1-s} + \left(\frac{1}{2}+t\right) \left(1-s\right) + 1 \right],
\end{align}
so that we finally arrive at~\cref{eq:Langevin_m>0k>0_benchmark_yy}
\begin{equation}
  \left\langle x(t)y(0) \right\rangle = \frac{r\left(A^+-A^-\right)}{8s^2(1-s)^{3/2}}  = \left\langle y(-t)x(0) \right\rangle.
\end{equation}
\end{proof}

\subsection{Time correlation function $\left\langle y(t)x(0) \right\rangle$}
\label{app:m>ok>o_ytx0}

\begin{proof}Here we need another intermediate result,
\begin{align}
  \left\langle \eta_y(t_1)G_y(t_2,s') \right\rangle
  &= \left\langle \int_{-\infty}^{t_2} e^{-s'(t_2-t')} \, \eta_y(t_1) \, \eta_y(t') \, \dd t' \right\rangle
  \nonumber \\
  &= \int_{-\infty}^{t_2} e^{-s'(t_2-t')}
  \left\langle \eta_y(t_1) \, \eta_y(t') \right\rangle \dd t'
  \nonumber \\
  &= \int_{-\infty}^{t_2} e^{-s'(t_2-t')} \, \delta(t_1-t') \, \dd t'
  \nonumber \\
  &= e^{-s(t_2-t_1)} \Theta(t_2-t_1) \,,
  \label{E.13}
\end{align}
as well as, with $y(t)$ from~\cref{eq:Langevin_m>0k>0_sol_y},
\begin{align}
  \left\langle \eta_y(t_1)y(t_2) \right\rangle
  &= \frac{1}{2\sqrt{1-s}} \left\langle \eta_y(t_1)\left[ G_y(t_2,s_-) - G_y(t_2,s_+) \right] \right\rangle
  \nonumber \\
  &= \frac{1}{2\sqrt{1-s}} \left[ \left\langle \eta_y(t_1)G_y(t_2,s_-) \right\rangle - \left\langle \eta_y(t_1)G_y(t_2,s_+) \right\rangle \right]
  \nonumber \\
  &= \frac{1}{2\sqrt{1-s}} \left[ e^{-s_-(t_2-t_1)} - e^{-s_+(t_2-t_1)} \right] \Theta(t_2-t_1) \,.
  \label{E.14}
\end{align}
Making use of~\cref{E.13,E.14}, one has
\begin{align}
  \left\langle G_y(t,a) G_x(0,b) \right\rangle
  &= \left\langle\int_{-\infty}^t e^{-a(t-t_1)} \, \eta_y(t_1) \, \dd t_1
  \int_{-\infty}^0 e^{-b(0-t_2)}\left[ 2ry(t_2) + \eta_x(t_2) \right] \dd t_2 \right\rangle
  \nonumber \\
  &= 2r \int_{-\infty}^t \int_{-\infty}^0 e^{-a(t-t_1)+bt_2}
  \left\langle \eta_y(t_1)y(t_2) \right\rangle  \dd t_2 \, \dd t_1
  \nonumber \\
  &= \frac{r}{\sqrt{1-s}} \int_{-\infty}^t \int_{-\infty}^0 e^{-a(t-t_1)+bt_2} \left[ e^{-s_-(t_2-t_1)} - e^{-s_+(t_2-t_1)} \right] \Theta(t_2-t_1) \dd t_2 \dd t_1
  \nonumber \\
  &= \frac{re^{-at}}{\sqrt{1-s}} \int_{-\infty}^{t_2} \int_{-\infty}^0 e^{at_1+bt_2} \left[ e^{-s_-(t_2-t_1)} - e^{-s_+(t_2-t_1)} \right]  \dd t_2 \, \dd t_1
  \nonumber \\
  &= \frac{re^{-at}}{\sqrt{1-s}} \left[ \frac{1}{a+b} \left( \frac{1}{a+s_-} - \frac{1}{a+s_+} \right) \right].
\end{align}
%
%
%
Starting from~\cref{eq:Langevin_m>0k>0_sol_x,eq:Langevin_m>0k>0_sol_y}, we can then immediately write down
\begin{align}
  \left\langle y(t)x(0) \right\rangle
  &= \frac{1}{4(1-s)} \left\langle \left[ G_y(t,s_-) - G_y(t,s_+) \right] \left[ G_x(0,s_-) - G_x(0,s_+) \right] \right\rangle
  \nonumber \\
  &= \frac{re^{-s_-t}}{16\sqrt{1-s}} \left( \frac{1}{s_-} \right)^2 - \frac{re^{-s_+t}}{16\sqrt{1-s}} \left( \frac{1}{s_+} \right)^2 \,.
\end{align}
Multiplying $\left(1-\sqrt{1-s}\right)^2\left(1+\sqrt{1-s}\right)^2=s^2$ on both sides gives
\begin{equation}
  s^2 \left\langle y(t)x(0) \right\rangle = \frac{re^{-s_-t}}{16\sqrt{1-s}} \left(1+\sqrt{1-s}\right)^2 - \frac{re^{-s_+t}}{16\sqrt{1-s}} \left(1-\sqrt{1-s}\right)^2 \,.
\end{equation}
so that we have proven~\cref{eq:Langevin_m>0k>0_benchmark_yx}
\begin{equation}
  \left\langle y(t)x(0) \right\rangle = \frac{r\left(C^-_2-C^+_2\right)}{16s^2\sqrt{1-s}}  = \left\langle x(-t)y(0) \right\rangle.
\end{equation}
\end{proof}

\subsection{Time correlation function $\left\langle x(t)x(0) \right\rangle$}
\label{app:m>ok>o_xtx0}

\begin{proof}For the sake of completeness and readers' convenience we here provide the full proof of~\cref{eq:Langevin_m>0k>0_benchmark_xx}. We begin, as before, with an intermediate result,
\begin{align}
  \left\langle G_y(t_1,a) G_y(t_2,b) \right\rangle
  &= \left\langle\int_{-\infty}^{t_1} e^{-a(t_1-t'_1)} \, \eta_y(t'_1) \, \dd t'_1
  \int_{-\infty}^{t_2} e^{-b(t_2-t'_2)} \, \eta_y(t'_2) \, \dd t'_2 \right\rangle
  \nonumber \\
  &= \int_{-\infty}^{t_1} \int_{-\infty}^{t_2} e^{-a(t_1-t'_1)-b(t_2-t'_2)} \, \dd t'_2 \, \dd t'_1
  \left\langle \eta_y(t'_1) \, \eta_y(t'_2) \right\rangle
  \nonumber \\
  &= \int_{-\infty}^{t_1} \int_{-\infty}^{t_2} e^{-a(t_1-t'_1)-b(t_2-t'_2)} \, \delta(t'_1-t'_2) \, \dd t'_2 \, \dd t'_1
  \nonumber \\
  &= e^{-(at_1+bt_2)} \int_{-\infty}^{\min(t_1,t_2)} e^{(a+b)t'_1} \, \dd t'_1
  \nonumber \\
  &= e^{-(at_1+bt_2)} \frac{e^{(a+b)\min(t_1,t_2)}}{a+b} \,, \quad \left[ \Re(a+b)>0 \right],
\end{align}
which corresponds, for $t_1\ge t_2$, or $t_1\le t_2$ to either
\begin{equation}
  \left\langle G_y(t_1,a) G_y(t_2,b) \right\rangle
  = e^{-(at_1+bt_2)} \frac{e^{(a+b)t_2}}{a+b} =  \frac{e^{-a(t_1-t_2)}}{a+b} \Theta(t_1-t_2)
  \label{E.20}
\end{equation}
or
\begin{equation}
  \left\langle G_y(t_1,a) G_y(t_2,b) \right\rangle
  = e^{-(at_1+bt_2)} \frac{e^{(a+b)t_1}}{a+b} =  \frac{e^{-b(t_2-t_1)}}{a+b} \Theta(t_2-t_1) \,.
  \label{E.21}
\end{equation}
With the help of~\cref{E.20,E.21,eq:Langevin_m>0k>0_sol_y}
\begin{align}
  \left\langle y(t_1)y(t_2) \right\rangle
  &= \frac{1}{4(1-s)} \left\langle \left[ G_y(t_1,s_-) - G_y(t_1,s_+) \right] \left[ G_y(t_2,s_-) - G_y(t_2,s_+) \right] \right\rangle  \nonumber \\
  &= \frac{1}{8s\sqrt{1-s}} \left[ s_+ e^{-s_-(t_1-t_2)} - s_- e^{-s_+(t_1-t_2)} \right] \Theta(t_1-t_2)
  \nonumber \\
  & \quad + \frac{1}{8s\sqrt{1-s}} \left[ s_+ e^{-s_-(t_2-t_1)} - s_- e^{-s_+(t_2-t_1)} \right] \Theta(t_2-t_1)
\,.
\end{align}
Defining $G_Y$ which differs from $G_y$ in that $\eta_y(t')$ is replaced by $y(t')$
\begin{equation}
  G_Y(t,s') \equiv \int_{-\infty}^t e^{-s'(t-t')} y(t') \, \dd t' \,,
\end{equation}
we have
\begin{align}
  \left\langle G_Y(t,a) G_Y(0,b) \right\rangle
  &= \left\langle\int_{-\infty}^t e^{-a(t-t_1)}y(t_1) \, \dd t_1
  \int_{-\infty}^0 e^{-b(0-t_2)}y(t_2) \, \dd t_2 \right\rangle
  \nonumber \\
  &= \int_{-\infty}^t \int_{-\infty}^0 e^{-a(t-t_1)+bt_2}
  \left\langle y(t_1)y(t_2) \right\rangle \dd t_2 \, \dd t_1
  \nonumber \\
  &= \frac{e^{-at}}{8s\sqrt{1-s}} \left[ s_+ \int_{-\infty}^0 e^{(b+s_-)t_2} \int_{t_2}^t e^{(a-s_-)t_1} \, \dd t_1 \, \dd t_2 \right.
  \nonumber \\
  & \quad\qquad\qquad\, \left. - s_- \int_{-\infty}^0 e^{(b+s_+)t_2} \int_{t_2}^t e^{(a-s_+)t_1} \, \dd t_1 \, \dd t_2 \right]
  \nonumber \\
  & \quad + \frac{e^{-at}}{8s\sqrt{1-s}} \left[ \frac{1}{a+b} \left( \frac{s_+}{a+s_-} - \frac{s_-}{a+s_+} \right) \right].
\end{align}
More specifically, the cases we really need below are
\begin{align*}
  \left\langle G_Y(t,s_-) G_Y(0,s_-) \right\rangle &= \frac{e^{-s_-t}}{8s\sqrt{1-s}} \left[ \frac{s_+t}{2s_-}+ \frac{s_+}{4s^2_-} -\frac{s_-}{s_--s_+} \left( \frac{e^{(s_--s_+)t}}{s_-+s_+} - \!\frac{1}{2s_-} \right) + \frac{b_+}{2s_-}\right],
  \nonumber \\
  \left\langle G_Y(t,s_-) G_Y(0,s_+) \right\rangle &= \frac{e^{-s_-t}}{8s\sqrt{1-s}} \left[ \frac{s_+t}{s_-+s_+} \!+\! \frac{s_+}{(s_-+s_+)^2}\! -\! \frac{s_-}{s_--s_+} \left( \frac{e^{(s_--s_+)t}}{2s_+} \!-\! \frac{1}{s_-+s_+} \right) \right]
  \nonumber \\
  & \quad + \frac{e^{-s_-t}}{8s\sqrt{1-s}} \left[ \frac{1}{s_-+s_+} \left( \frac{s_+}{2s_-} - \frac{s_-}{s_-+s_+} \right) \right], \\
  \left\langle G_Y(t,s_+) G_Y(0,s_-) \right\rangle &= \frac{e^{-s_+t}}{8s\sqrt{1-s}} \left[ \frac{s_+}{s_+-s_-} \left( \frac{e^{(s_+-s_-)t}}{2s_-} \!-\! \frac{1}{s_-+s_+} \right) \!-\! \frac{s_-t}{s_-+s_+} \!-\! \frac{s_-}{(s_-+s_+)^2} \right]
  \nonumber \\
  & \quad + \frac{e^{-s_+t}}{8s\sqrt{1-s}} \left[ \frac{1}{s_-+s_+} \left( \frac{s_+}{s_-+s_+} - \frac{s_-}{2s_+} \right) \right], \nonumber \\
  \left\langle G_Y(t,s_+) G_Y(0,s_+) \right\rangle &= \frac{e^{-s_+t}}{8s\sqrt{1-s}} \left[ \frac{s_+}{s_+-s_-} \left( \frac{e^{(s_+-s_-)t}}{s_-+s_+} - \frac{1}{2s_+} \right) - \frac{s_-t}{2s_+} - \frac{s_-}{4s^2_+} + \frac{b_-}{2s_+}  \right].
\end{align*}
where the abbreviation
\begin{equation}
 b_\pm = \pm \frac{s_\pm}{2s_\mp} \mp \frac{s_\mp}{s_-+s_+}
\end{equation}
was needed.
We can rewrite the solution~\cref{eq:Langevin_m>0k>0_sol_x} as the sum of two uncorrelated parts
$x(t) = x_1(t) + x_2(t)$, with
\begin{equation}\label{eq:Langevin_m>0k>0_sol_x1}
  x_i(t) = \frac{1}{2\sqrt{1-s}} \left[ G_{x_i}(t,s_-) - G_{x_i}(t,s_+) \right], \quad i=1,2 \,,
\end{equation}
and
\begin{equation}\label{eq:Langevin_m>0k>0_sol_Gx1x2}
  G_{x_1}(t,s') \equiv \int_{-\infty}^t e^{-s'(t-t')} 2ry(t') \, \dd t' \,, \quad G_{x_2}(t,s') \equiv \int_{-\infty}^t e^{-s'(t-t')} \eta_x(t') \, \dd t' \,.
\end{equation}
%
Since $\langle x_2(t)x_2(0)\rangle=\langle y(t)y(0)\rangle$ had already been calculated above,
the remaining contribution to $\langle x(t)x(0)\rangle$ is
\begin{align}
  \left\langle x_1(t)x_1(0) \right\rangle &= \frac{r^2}{1-s} \left\langle \left[ G_Y(t,s_-) - G_Y(t,s_+) \right] \left[ G_Y(0,s_-) - G_Y(0,s_+) \right] \right\rangle
  \nonumber \\
  &= \frac{r^2e^{-s_+t}}{16s(1-s)^{3/2}} \frac{\sqrt{1-s}}{\left(1+\sqrt{1-s}\right)} \left[ s_-t + \frac{s_-}{s_+} - \left( \sqrt{1-s} - \frac{1}{\sqrt{1-s}} \right) \right] \nonumber \\
  & \quad + \frac{r^2e^{-s_-t}}{16s(1-s)^{3/2}} \frac{\sqrt{1-s}}{\left(1-\sqrt{1-s}\right)} \left[ s_+t + \frac{s_+}{s_-} + \left( \sqrt{1-s} - \frac{1}{\sqrt{1-s}} \right) \right].
\end{align}
Multiplying $\left(1-\sqrt{1-s}\right)^2\left(1+\sqrt{1-s}\right)^2=s^2$ on both sides gives
\begin{align}
  s^2 \left\langle x_1(t)x_1(0) \right\rangle
  &= \frac{r^2e^{-s_+t}}{16s(1-s)^{3/2}} \left(1-\sqrt{1-s}\right)^2 \left[ \sqrt{1-s} \left( st + s + 1 \right) + 2s - 1 \right]
  \nonumber \\
  & \quad + \frac{r^2e^{-s_-t}}{16s(1-s)^{3/2}} \left(1+\sqrt{1-s}\right)^2 \left[ \sqrt{1-s} \left( st + s + 1 \right) - 2s + 1 \right].
\end{align}
which brings us in agreement with~\cref{eq:Langevin_m>0k>0_benchmark_xx}
\begin{equation}
  \left\langle x(t)x(0) \right\rangle = \frac{\left( C_1^+ + C_1^- \right)}{8 s\sqrt{1-s}}  + \frac{r^2\left( B^+ + B^- \right)}{16s^3(1-s)^{3/2}}  \,.
\end{equation}
\end{proof}

\section*{Acknowledgments}
The authors thank Hans Christian \"{O}ttinger for valuable suggestions and comments.

\bibliographystyle{siamplain}
\bibliography{refs-SIAM}

\end{document}

%% file: ShKr2018-shared-SIAM.tex
\usepackage{lipsum}
\usepackage{amsfonts}
\usepackage{graphicx}
\usepackage{epstopdf}
\usepackage{algorithmic}
\ifpdf
  \DeclareGraphicsExtensions{.eps,.pdf,.png,.jpg}
\else
  \DeclareGraphicsExtensions{.eps}
\fi

\usepackage{enumitem}
\setlist[enumerate]{leftmargin=.5in}
\setlist[itemize]{leftmargin=.5in}

\newsiamremark{remark}{Remark}
\newsiamremark{hypothesis}{Hypothesis}
\crefname{hypothesis}{Hypothesis}{Hypotheses}
\newsiamthm{claim}{Claim}

\headers{Langevin and Brownian dynamics: Analytic treatment and numerics}{X. Shang and M. Kr\"oger}

\title{Time correlation functions of equilibrium and nonequilibrium Langevin dynamics: Derivations and numerics using random numbers\thanks{Submitted to the editors \today.
}}

\author{Xiaocheng Shang \and Martin Kr\"oger\thanks{Polymer Physics, Department of Materials, ETH Zurich, Leopold-Ruzicka-Weg 4, CH-8093 Zurich, Switzerland
  (\email{x.shang@mat.ethz.ch}, \email{mk@mat.ethz.ch}, \url{http://www.complexfluids.ethz.ch}).}
}

\usepackage{amsopn}
